%% file: arxiv.tex
\pdfoutput=1
\documentclass[sigconf, screen, nonacm]{acmart}

\input{preamble}

\begin{document}

\title{\sys: Fast Dynamic Sparse Prefill on a Wafer-Scale Accelerator}

\author{Jianchang Su}
\affiliation{%
  \institution{University of Connecticut}
  \city{Storrs}
  \state{Connecticut}
  \country{USA}}
\email{jianchang.su@uconn.edu}

\author{Yifan Zhang}
\affiliation{%
  \institution{University of Connecticut}
  \city{Storrs}
  \state{Connecticut}
  \country{USA}}
\email{yifan.4.zhang@uconn.edu}

\author{Wei Zhang}
\affiliation{%
  \institution{University of Connecticut}
  \city{Storrs}
  \state{Connecticut}
  \country{USA}}
\email{wei.13.zhang@uconn.edu}

\input{sections/00_abstract}

\maketitle

\input{sections/01_introduction}
\input{sections/02_background}
\input{sections/03_motivation}
\input{sections/04_design}
\input{sections/05_implementation}
\input{sections/06_evaluation}
\input{sections/07_discussion}
\input{sections/08_related}
\input{sections/09_conclusion}
\input{sections/99_ack}

\bibliographystyle{ACM-Reference-Format}
\bibliography{references}

\end{document}

%% file: preamble.tex
\usepackage{microtype}
\usepackage{xspace}
\usepackage{enumitem}

\expandafter\def\expandafter\UrlBreaks\expandafter{\UrlBreaks
  \do\a\do\b\do\c\do\d\do\e\do\f\do\g\do\h\do\i\do\j\do\k\do\l\do\m\do\n\do\o\do\p\do\q
  \do\r\do\s\do\t\do\u\do\v\do\w\do\x\do\y\do\z\do\A\do\B\do\C\do\D\do\E\do\F\do\G\do\H
  \do\I\do\J\do\K\do\L\do\M\do\N\do\O\do\P\do\Q\do\R\do\S\do\T\do\U\do\V\do\W\do\X\do\Y
  \do\Z\do0\do1\do2\do3\do4\do5\do6\do7\do8\do9}

\input{figstyle}
\input{figures/style}

\makeatletter
\renewcommand*\sectionautorefname{\S\@gobble}
\renewcommand*\subsectionautorefname{\S\@gobble}
\renewcommand*\subsubsectionautorefname{\S\@gobble}
\makeatother

\newcommand{\sys}{SparseWafer\xspace}

\newcommand{\takeaway}[1]{%
  \par\addvspace{3pt}\noindent
  {\setlength{\fboxsep}{2pt}\setlength{\fboxrule}{0.3pt}%
   \fbox{\parbox{\dimexpr\linewidth-2\fboxsep-2\fboxrule\relax}{%
     \textbf{Takeaway.} #1}}}%
  \par\addvspace{2pt}}


%% file: figstyle.tex
\usepackage{xcolor}
\usepackage{tikz}
\usetikzlibrary{positioning,arrows.meta,fit,backgrounds,calc,patterns,
  decorations.pathreplacing}
\usepackage[caption=false,labelformat=simple,font=footnotesize,captionskip=1pt,farskip=2pt,
  nearskip=0pt]{subfig}

\definecolor{cLine}{HTML}{3A3A3A}   
\definecolor{cDark}{HTML}{262626}   

\tikzset{
  >={Stealth[length=3.5pt]},
  region/.style={draw=cLine!80, fill=black!5, rounded corners=3pt, line width=0.5pt},
  part/.style={draw=cLine!60, fill=white, rounded corners=1.5pt, line width=0.4pt,
    align=center, font=\tiny, inner sep=2pt},
  new/.style={draw=cDark, fill=black!15, rounded corners=1.5pt, line width=0.8pt,
    align=center, font=\tiny\bfseries, inner sep=2pt},
  tok/.style={draw=cLine!70, fill=white, rounded corners=1pt, inner sep=1.5pt,
    minimum height=0.34cm, font=\tiny\ttfamily, line width=0.4pt},
  computed/.style={tok, draw=cDark, fill=cDark, text=white},
  guess/.style={tok, draw=cLine!60, fill=black!4, dashed},
  flow/.style={->, line width=0.6pt, cLine},
  newflow/.style={->, line width=1pt, cDark},
  busy/.style={fill=black!35, draw=cLine, line width=0.4pt},
  host/.style={fill=black!8, draw=cLine!60, line width=0.4pt},
  work/.style={fill=black!20, draw=cDark, line width=0.6pt},
  idle/.style={pattern=north east lines, pattern color=cLine!60, draw=cLine!70,
    line width=0.4pt},
  num/.style={circle, fill=cDark, text=white, font=\tiny\bfseries, inner sep=0.5pt},
}

%% file: figures/style.tex
\definecolor{sysblue}{HTML}{2A78D6}
\definecolor{sysbluedark}{HTML}{1C5CAB}
\definecolor{sysbluelight}{HTML}{DCE9F9}
\definecolor{systeal}{HTML}{199E70}
\definecolor{systeallight}{HTML}{D8F1E7}
\definecolor{sysviolet}{HTML}{6250D6}
\definecolor{sysmagenta}{HTML}{D55181}
\definecolor{sysink}{HTML}{2E2E2E}
\definecolor{sysline}{HTML}{6E6E6E}
\definecolor{sysfill}{HTML}{F3F3F3}
\definecolor{sysgrid}{HTML}{E2E2E2}
\newcommand{\figsmall}{\fontsize{6.2}{7.2}\selectfont}
\colorlet{unitA}{sysblue!60}
\colorlet{unitB}{systeal!60}
\colorlet{unitC}{sysviolet!55}
\colorlet{unitD}{sysmagenta!55}
\tikzset{
  sys/.style={font=\sffamily\scriptsize, line join=round, text=sysink,
              >={Latex[length=1.5mm,width=1.1mm]}},
  pe/.style={draw=sysline, fill=white, minimum size=3.2mm, inner sep=0pt, line width=0.4pt},
  relay/.style={draw=systeal, fill=systeallight, minimum size=3.2mm, inner sep=0pt,
                line width=0.5pt},
  store/.style={draw=systeal, fill=systeal!45, minimum size=3.2mm, inner sep=0pt,
                line width=0.5pt},
  box/.style={draw=sysline, rounded corners=1.5pt, fill=sysfill, line width=0.5pt},
  ourbox/.style={draw=sysbluedark, rounded corners=1.5pt, fill=sysbluelight, line width=0.7pt},
  kv/.style={->, systeal, line width=0.8pt},
  flow/.style={->, sysink, line width=0.9pt},
  ctrl/.style={->, sysbluedark, line width=0.7pt, densely dashed},
  num/.style={circle, fill=sysink, text=white, inner sep=0pt, minimum size=3.6mm,
              font=\sffamily\bfseries\figsmall},
  lbl/.style={align=center, inner sep=1pt, text=sysink},
  note/.style={align=center, inner sep=1pt, text=sysline, font=\sffamily\figsmall},
  head/.style={font=\sffamily\bfseries\scriptsize, text=sysink, inner sep=1pt},
  stall/.style={pattern=north east lines, pattern color=sysline!80},
}
\DeclareRobustCommand{\step}[1]{\tikz[baseline=(s.base)]\node[num, minimum size=3.1mm] (s) {#1};}
\newcommand{\unitpe}[3]{%
  \node[pe] (#1) at (#2) {};
  \foreach \u [count=\i from 0] in {#3} {
    \fill[\u] ($(#1.north west)+(0.032,-0.032-0.072*\i)$) rectangle
              ($(#1.north east)+(-0.032,-0.092-0.072*\i)$);
  }%
}

%% file: sections/00_abstract.tex
\begin{abstract}
Dynamic block-sparse attention skips most of the attention work of a long prompt. On
accelerators whose multicast is statically routed and flow-controlled, such as
wafer-scale engines, the pruning gains far less than the skipped work suggests: with the
conventional placement, sparse attention that skips most of the blocks still takes most
of the dense attention time. The cause is multicast backpressure: a route advances only when every destination has a
free buffer, so processing elements (PEs) that receive the same key-value stream stall
each other when their work falls at different points of it. We present \sys{}, which
reduces this stall by placing query rows. The rows that share one selection decision
select the same key blocks, so spreading them cyclically over the PEs of a stream gives
every PE the same work on every block, and we prove that such PEs run as fast as one PE
alone, for any buffer depth. A per-head start offset, which starts each query head at a
different set of PEs, extends this balance across the multicast routes that share one
key-value stream; it depends only on tensor shapes and is fixed before run time. We evaluate \sys{} on the Cerebras SDK's WSE-3 simulator and
a model calibrated on it, with selection and data movement included. With the masks of a
published selection method (XAttention), \sys{} is 1.31--1.49$\times$ faster than sparse
prefill with the conventional placement at each model's longest context. This raises
the speedup of sparse prefill over dense prefill from 1.13--1.35$\times$ to
1.47--1.90$\times$. With these masks, sparse prefill loses at most 1.8 points of
accuracy on the RULER long-context benchmark, and on the denser masks of RULER prompts
\sys{} is still 1.28--1.54$\times$ faster than dense prefill at 128K. It is also faster
than mask-aware load balancers and within 5--10\% in attention time of an idealized,
zero-cost one: on such networks, the shape of the computation decides almost all of
where each row belongs.
\end{abstract}

%% file: sections/01_introduction.tex
\section{Introduction}
\label{sec:intro}

Prefill, the forward pass over a prompt before the first output token, sets the time
to first token (TTFT) of long-context inference. Its attention grows with the square of
the context length: at 128K tokens, one causal layer of Llama-3.1-8B evaluates
$2.7\times10^{11}$ query-key pairs. Dynamic sparse attention reduces this work. It
first estimates which key blocks carry most of each query block's softmax probability,
then computes only those blocks. Published methods differ in how they estimate, for example
by strided antidiagonal sums~\cite{xu2025xattention} or by vertical and diagonal key
lines~\cite{jiang2024minference}, and on GPUs they keep their accuracy at a fraction of
the work.

Spatial accelerators are a different target for the same computation. Like other
spatial machines~\cite{prabhakar2024sambanova,jia2019dissecting,talpes2023microarchitecture},
a Cerebras WSE-3 keeps data in SRAM next to its nearly 900,000 processing elements
(PEs); unlike a GPU, it multicasts blocks over statically configured, flow-controlled
routes of its on-chip network (Cerebras'
fabric)~\cite{lie2023cerebras,cerebras2026cerebras}.
WaferLLM has shown that dense LLM inference maps onto such a wafer~\cite{he2025waferllm}.
We target dynamic sparse prefill on such a network, with every cost of the sparse
computation included. We study it on the WSE-3; the multicast stall that we describe
next arises on any network whose multicast route advances only when every destination
can accept data.

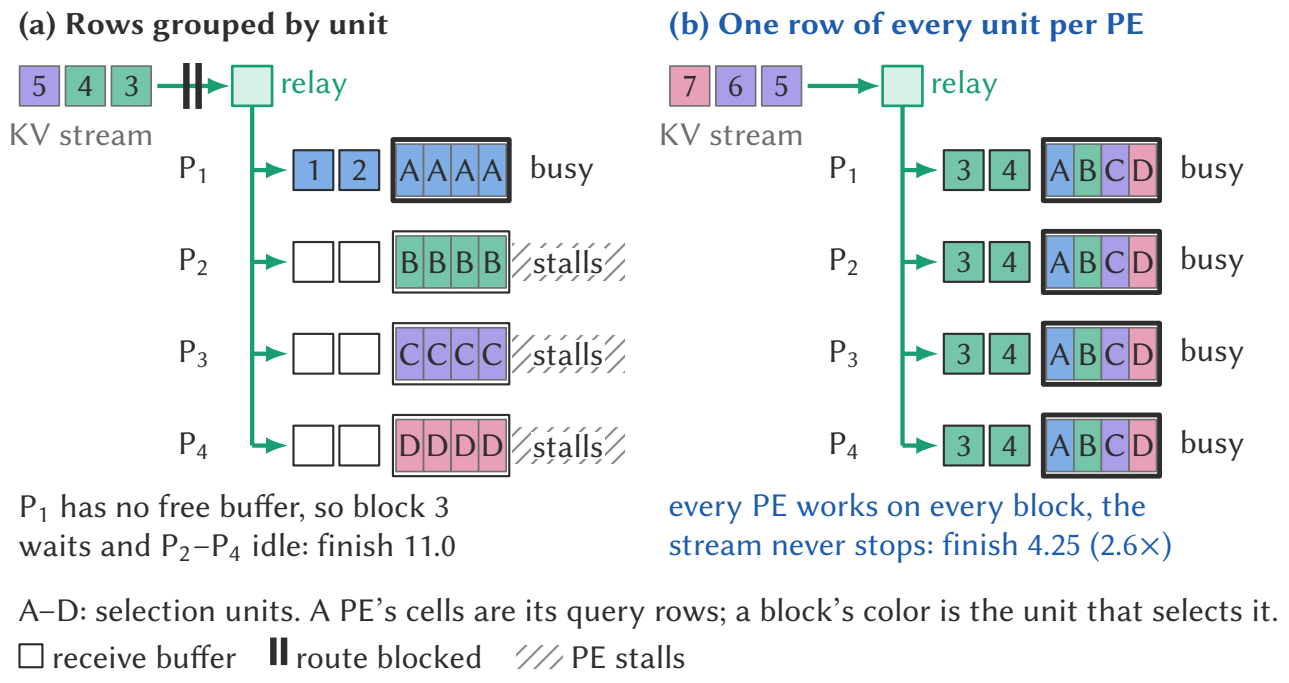
\begin{figure}[t]
  \centering
  \resizebox{\columnwidth}{!}{\input{figures/intro}}
  \caption{The placement of query rows decides whether the PEs that share a KV stream
  stall each other. One head's KV blocks reach four PEs with two receive buffers each,
  and the route advances only when every PE has a free buffer; units A--D select blocks
  1--2, 3--4, 5--6, 7--8. (a)~Each PE holds whole units (contiguous placement);
  P\textsubscript{1} still works on blocks 1--2, so block 3 cannot enter. (b)~All PEs work on every block and finish 2.6$\times$ sooner
  (\autoref{eq:stream}).}
  \label{fig:intro}
\end{figure}

We observe that on a statically routed wafer, the reduction in attention work does not
give a proportional speedup. At 128K tokens, the masks of one such method,
XAttention~\cite{xu2025xattention}, select
11--26\% of the causal blocks. Yet with the conventional striped
placement~\cite{brandon2023striped}, which gives a head's consecutive tokens to
consecutive PEs, sparse prefill is only 1.02--1.35$\times$ faster than dense
prefill (\autoref{sec:motivation}).
The main cause is multicast backpressure. Key and value (KV) blocks travel over multicast
routes to the PEs that hold query rows. A route forwards a block only when every
destination has a free receive buffer. The PEs that share a route therefore stall each
other wherever their selected blocks differ along the stream (\autoref{fig:intro}(a)).
With striped placement, this stall arises between groups that share one KV stream, and
the PEs also have unequal total work; a load balancer that equalizes total
work still stalls on the stream (\autoref{fig:column-parts}).

Existing placements do not prevent this stall. Placements for distributed attention
balance the total work of every processor~\cite{brandon2023striped,nvidia2026context},
and partitioners for sparse workloads also reduce the communication
volume~\cite{catalyurek1999hypergraph,sun2026sparsp}. The stall, however, depends on
where each PE's work falls along the stream, not on how much total work it has. Quartz
partitions sparse data so that all tiles stay busy at every moment~\cite{golden2025quartz},
but it computes the partition from the data before the program runs. Dynamic sparse
attention produces its masks only after selection, so such a partition would have to
run on the same PEs, on the critical path of every layer.

Our key insight is that the query rows that share one selection decision select the same
key blocks. We call these rows a \emph{selection unit}. On the wafer, a \emph{consumer}
is a set of $F$ vertically adjacent PEs that hold the same query rows; consumers form
multicast groups of 64, and the groups of one KV head share one stream, sent down one
column of PEs (the column stream). If every
consumer of a group holds the same number of rows of every unit, all of them have the
same work on every block of the stream. We prove that the group then runs as fast as one
consumer alone, for any buffer depth and delivery time (\autoref{prop:equal}). Striped
placement already meets this condition inside a group, because
a group's consumers hold consecutive tokens of the same units; its stall lies between
the groups that share a column stream.

Applying this insight raises three challenges. First, striped placement gives each group
the same token positions of every head, so the groups that share a column stream stall
each other. Second, KV heads select different numbers of blocks, so the columns of a
layer finish at different times. Third, the selection estimate runs on the critical
path of every layer, and at short contexts it costs more than it saves.

To address these challenges, we propose \sys{} (\autoref{sec:design}). First, \sys{}
keeps each unit's rows cyclic over the consumers of a group~\cite{blackford1997scalapack},
as striped placement does, but starts each query head at a different group of the KV
column, so the cumulative work of every group stays close to the column's mean. This \emph{unit-cyclic placement} depends only on tensor shapes, so
it is fixed before the prompt arrives; it makes end-to-end prefill 1.16--1.45$\times$
faster than striped placement with the same masks at 128K. Second, \sys{} interleaves
KV heads over the KV columns, which gives a further speedup of up to 1.19$\times$ on
Llama-3.1-8B (\autoref{sec:eval-breakdown}). Third, it runs block selection on the attention PEs with
a fused scoring loop, which shortens the \emph{break-even length}, the shortest
context at which sparse prefill is faster than dense prefill, by 8--14\%.

We implement \sys{} on the Cerebras SDK and run it on the SDK's WSE-3 simulator, which
reports cycles per PE. We then estimate end-to-end prefill with a performance
model calibrated on those measurements (\S\S\ref{sec:impl} and~\ref{sec:eval-setup}). On
three current LLMs and two published selection methods, with selection, the query
permutation and expert routing included, \sys{} with XAttention masks is
1.31--1.49$\times$ faster than striped sparse prefill at each model's longest context
(128K or 256K tokens). It thus raises the speedup of sparse prefill over dense prefill
from 1.13--1.35$\times$ to 1.47--1.90$\times$, and with these masks sparse prefill loses
at most 1.8 points of RULER accuracy. \sys{} breaks even with dense prefill at contexts
1.4--2.7$\times$ shorter than striped placement does, and it is faster than mask-aware
load balancers, even at zero decision cost.

Our contributions are as follows:
\begin{itemize}
\item We characterize dynamic sparse prefill on a statically routed wafer and show that
  multicast backpressure between PEs that share a stream, rather than data volume or
  total work, limits its speedup (\autoref{sec:motivation}).
\item We propose unit-cyclic placement, which distributes the rows of every selection
  unit cyclically over the consumers of a group and offsets each query head's start
  across the groups of a KV column, and prove with a max-plus model that equal work per
  block removes the stall inside a group for any buffer depth
  (\S\S\ref{sec:design-group}--\ref{sec:design-column}).
\item We design and implement \sys{} on the WSE-3 with block selection on the attention
  PEs, one query permutation, head interleaving and an offline planner
  (\S\S\ref{sec:design} and~\ref{sec:impl}).
\item We evaluate \sys{} on three LLMs against conventional placements, mask-aware load
  balancers and dense prefill, with every cost of sparse prefill included, and measure
  its accuracy on RULER (\autoref{sec:eval}).
\end{itemize}

%% file: figures/intro.tex
\begin{tikzpicture}[sys, x=1cm, y=1cm]
  \def\bs{0.26}
  \newcommand{\blk}[4]{
    \node[draw=sysline, fill=#3, minimum size=\bs cm, inner sep=0pt, line width=0.3pt,
          font=\sffamily\figsmall] at (#1,#2) {#4};
  }
  \newcommand{\buf}[4]{
    \node[draw=sysink, fill=#3, minimum size=\bs cm, inner sep=0pt, line width=0.4pt,
          font=\sffamily\figsmall] at (#1,#2) {#4};
  }
  \newcommand{\rows}[4]{
    \foreach \cellfill/\cellname [count=\cellindex from 0] in {#4} {
      \node[draw=sysline, fill=\cellfill, line width=0.3pt, minimum width=0.18cm,
            minimum height=0.34cm, inner sep=0pt, font=\sffamily\figsmall, text=sysink]
        at ({#1-0.27+0.18*\cellindex},#2) {\cellname};
    }
    \draw[sysink, line width=#3] ({#1-0.38},{#2-0.2}) rectangle ({#1+0.38},{#2+0.2});
  }
  \newcommand{\route}[1]{
    \node[note, anchor=north] at ({#1+0.4},-0.2) {KV stream};
    \draw[kv] ({#1+0.9},0) -- ({#1+1.38},0);
    \node[relay, minimum size=2.6mm] at ({#1+1.52},0) {};
    \node[note, anchor=west, text=systeal] at ({#1+1.66},0) {relay};
    \draw[kv, -] ({#1+1.52},-0.13) -- ({#1+1.52},-2.35);
    \foreach \i in {1,...,4} {
      \pgfmathsetmacro{\yy}{-0.55-0.6*(\i-1)}
      \draw[kv] ({#1+1.52},\yy) -- ({#1+1.76},\yy);
      \node[font=\sffamily\figsmall, text=sysink, anchor=east] at ({#1+1.36},\yy) {P\textsubscript{\i}};
    }
  }
  \def\ox{0}
  \node[anchor=south west, inner sep=1pt, lbl, align=left] at (\ox-0.05,0.25)
    {\textbf{(a) Rows grouped by unit}};
  \blk{\ox+0.13}{0}{unitC}{5}\blk{\ox+0.43}{0}{unitB}{4}\blk{\ox+0.73}{0}{unitB}{3}
  \route{\ox}
  \draw[sysink, line width=1.2pt] ({\ox+1.09},-0.15) -- ({\ox+1.09},0.15);
  \draw[sysink, line width=1.2pt] ({\ox+1.17},-0.15) -- ({\ox+1.17},0.15);
  \buf{\ox+1.92}{-0.55}{unitA}{1}\buf{\ox+2.22}{-0.55}{unitA}{2}
  \rows{\ox+2.82}{-0.55}{0.9pt}{unitA/A,unitA/A,unitA/A,unitA/A}
  \node[font=\sffamily\figsmall, text=sysink, anchor=west] at (\ox+3.22,-0.55) {busy};
  \foreach \yy in {-1.15,-1.75,-2.35} {
    \buf{\ox+1.92}{\yy}{white}{}\buf{\ox+2.22}{\yy}{white}{}
    \fill[stall] ({\ox+3.22},{\yy-0.12}) rectangle ({\ox+3.95},{\yy+0.12});
    \node[font=\sffamily\figsmall, text=sysink, fill=white, inner sep=0.3pt] at (\ox+3.585,\yy) {stalls};
  }
  \rows{\ox+2.82}{-1.15}{0.4pt}{unitB/B,unitB/B,unitB/B,unitB/B}
  \rows{\ox+2.82}{-1.75}{0.4pt}{unitC/C,unitC/C,unitC/C,unitC/C}
  \rows{\ox+2.82}{-2.35}{0.4pt}{unitD/D,unitD/D,unitD/D,unitD/D}
  \node[note, anchor=north west, align=left, text=sysink] at (\ox-0.05,-2.62)
    {P\textsubscript{1} has no free buffer, so block 3\\waits and P\textsubscript{2}--P\textsubscript{4} idle: finish 11.0};
  \def\ox{4.25}
  \node[anchor=south west, inner sep=1pt, lbl, align=left, text=sysbluedark] at (\ox-0.05,0.25)
    {\textbf{(b) One row of every unit per PE}};
  \blk{\ox+0.13}{0}{unitD}{7}\blk{\ox+0.43}{0}{unitC}{6}\blk{\ox+0.73}{0}{unitC}{5}
  \route{\ox}
  \foreach \i in {1,...,4} {
    \pgfmathsetmacro{\yy}{-0.55-0.6*(\i-1)}
    \buf{\ox+1.92}{\yy}{unitB}{3}\buf{\ox+2.22}{\yy}{unitB}{4}
    \rows{\ox+2.82}{\yy}{0.9pt}{unitA/A,unitB/B,unitC/C,unitD/D}
    \node[font=\sffamily\figsmall, text=sysink, anchor=west] at (\ox+3.22,\yy) {busy};
  }
  \node[note, anchor=north west, align=left, text=sysbluedark] at (\ox-0.05,-2.62)
    {every PE works on every block, the\\stream never stops: finish 4.25 (2.6$\times$)};
  \node[anchor=north west, font=\sffamily\figsmall, text=sysink, inner sep=1pt,
        align=left] at (-0.05,-3.3)
    {A--D: selection units. A PE's cells are its query rows; a block's color is the unit
     that selects it.\\
     \tikz\draw[sysink, line width=0.4pt] (0,0) rectangle (0.15,0.15);\ receive buffer\quad
     \tikz{\draw[sysink, line width=1.2pt] (0,0) -- (0,0.18);
           \draw[sysink, line width=1.2pt] (0.07,0) -- (0.07,0.18);}\ route blocked\quad
     \tikz\fill[stall] (0,0) rectangle (0.3,0.15);\ PE stalls};
\end{tikzpicture}%

%% file: sections/02_background.tex
\section{Background}
\label{sec:background}

\subsection{Dynamic sparse prefill}

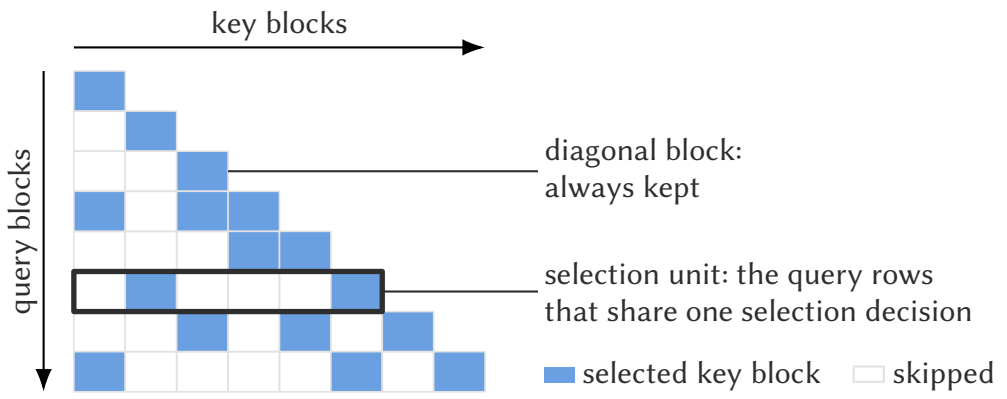
\begin{figure}[t]
  \centering
  \input{figures/masks}
  \caption{A dynamic block-sparse mask of one head (schematic). Each selection unit, here
  one query block, keeps a subset of the earlier key blocks and the diagonal block.}
  \label{fig:masks}
\end{figure}

Causal attention splits the queries and keys of a head into blocks. Block-sparse
prefill computes, for each query block, only a selected subset of the causally
preceding key blocks, always including the diagonal block (\autoref{fig:masks}).
Dynamic methods select at run time, per input, in two steps: a low-cost \emph{estimate} of
each key block's share of the softmax probability, a block \emph{score}, then a
\emph{selection} rule over the scores, such as a threshold or a top-$k$. The result
is one mask of key blocks per \emph{selection unit}, the query rows that share one
selection decision. All rows of a unit use the same key blocks, except for causal
masking inside the diagonal block. We call the fraction of the causal blocks that a
mask selects its \emph{density} (measured in \autoref{sec:motivation}).

Methods differ in their estimate and in the shape of their units. Block methods select
per query block and query head; MInference~\cite{jiang2024minference} builds a pattern
of vertical key columns and diagonal (slash) lines per head, which its kernel processes
in 64-token query blocks; per-token methods such as NSA~\cite{yuan2025native} and
DeepSeek sparse attention~\cite{deepseekai2025deepseek} share one selection among the
query heads of a KV head under grouped-query attention (GQA). We evaluate the two
published block-sparse prefill methods that recent prefill work compares against, at
their published settings. XAttention~\cite{xu2025xattention} reshapes queries and keys
with stride $S=8$, so each reshaped score sums an antidiagonal of an $8\times8$ tile,
and keeps for each 128-token query block of a head the smallest set of 128-token key
blocks whose scores reach a threshold share of the total (a published threshold per head
for Llama-3.1-8B; 0.95 for the Qwen3 models). MInference estimates from the last 64 queries of each head. The
selection unit is the structure that the placement in \autoref{sec:design} exploits.

\subsection{A wafer-scale accelerator}

\begin{figure}[t]
  \centering
  \resizebox{\columnwidth}{!}{\input{figures/wafer}}
  \caption{A multicast route advances at the rate of its slowest destination, as the SDK
  simulator runs it. (a) The PE mesh. (b) One PE.}
  \label{fig:wafer}
\end{figure}
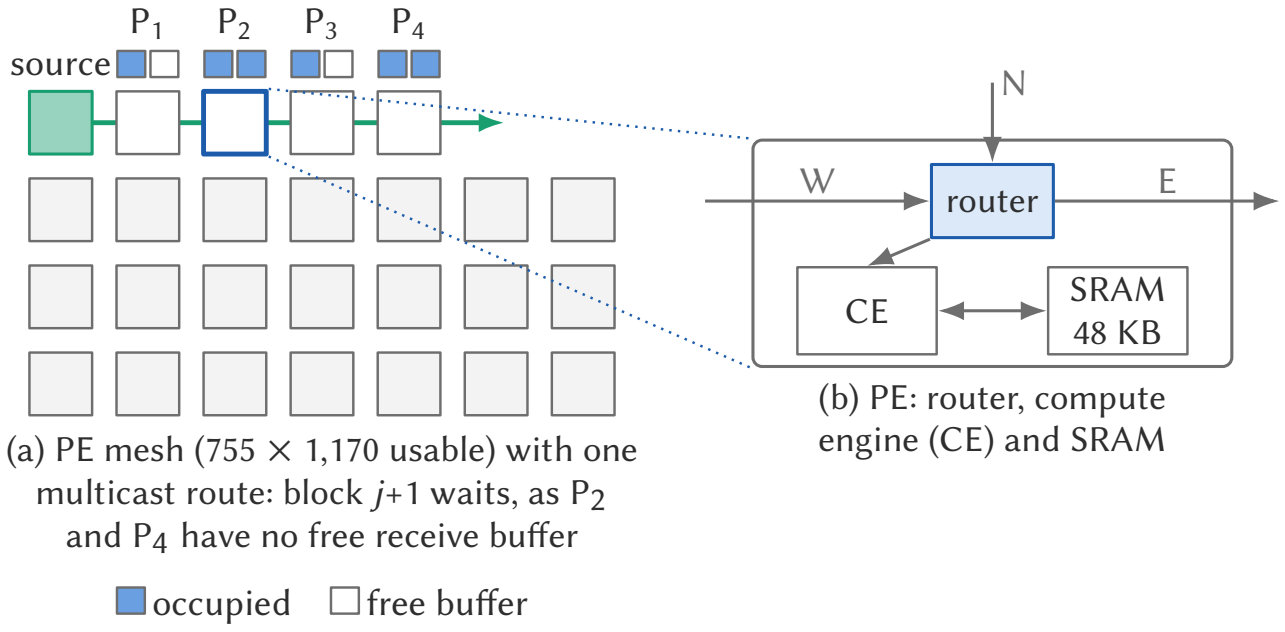

The Cerebras WSE-3 is a 2D mesh of PEs on one wafer (\autoref{fig:wafer}). Each PE has a
router, a compute engine (CE) with FP16 vector multiply-add instructions, and
48\,KB of SRAM for code, data and receive buffers~\cite{cerebras2026cerebras}. The router has
links to its four neighbors and a local port to its own CE (Cerebras'
ramp)~\cite{cerebras2026wafer}. A program built with the SDK can use
$755\times1{,}170$ PEs~\cite{cerebras2026cerebras}. Data moves in 32-bit packets
(wavelets), each on one of the router's virtual channels (colors). For each virtual
channel, a route configured before the program runs names the ports a packet leaves
by~\cite{cerebras2026wafer}, so one route can deliver each packet to every PE along a
path (multicast). The on-chip network is flow-controlled~\cite{lie2023cerebras}. On the
SDK simulator, a multicast route forwards a packet only when every destination can
accept it, so a PE whose receive buffers are
full blocks the route for all its destinations. This is
\emph{backpressure}~\cite{dally2004principles}.

\subsection{Attention on the wafer}
\label{sec:mapping}

We follow the dataflow of spatial attention mappings: query rows stay on PEs, and keys
and values stream past them. We call the sequence of K and V blocks that one multicast
route delivers a \emph{stream}; the PEs on the route share it. For each KV head, the
\emph{KV source}, the PEs that hold the K and V blocks that the projection produces,
sends them down one column of PEs, the head's \emph{KV column}; we call this the
\emph{column stream} (\autoref{fig:overview}). Each head's 128 dimensions are split over
$F$ PEs ($F=4$ or 8), so a \emph{consumer} is a set of $F$ vertically adjacent PEs that
hold the same query rows. A \emph{relay} PE in each PE row of the column receives the
column stream on its local port. It forwards along its PE row, to a \emph{multicast group} of
$G=64$ consumers, only the blocks that at least one query row of the group selected.
For each block it receives, a consumer computes the rows that selected the block, with
an online softmax~\cite{milakov2018online} as in
FlashAttention~\cite{dao2022flashattention}. Each PE of a consumer has two receive
buffers. Dense prefill uses the same mapping with every causal block.

%% file: figures/masks.tex
\begin{tikzpicture}[sys, x=1cm, y=0.78cm]
  \def\s{0.34}
  \tikzset{cell/.style={draw=sysgrid, line width=0.3pt},
           lead/.style={sysink, line width=0.4pt}}
  \foreach \r in {0,...,7} {
    \foreach \c in {0,...,7} {
      \pgfmathtruncatemacro{\causal}{\c<=\r ? 1 : 0}
      \ifnum\causal=1
        \pgfmathtruncatemacro{\sel}{(\c==\r || mod(\r*5+\c*3,7)<2) ? 1 : 0}
        \ifnum\sel=1
          \fill[sysblue!70] (\s*\c, -\s*\r) rectangle ++(\s, -\s);
        \fi
        \draw[cell] (\s*\c, -\s*\r) rectangle ++(\s, -\s);
      \fi
    }
  }
  \draw[sysink, line width=1pt] (0,-\s*5) rectangle (\s*6,-\s*6);
  \draw[->] (0,0.2) -- node[above, lbl] {key blocks} (\s*8,0.2);
  \draw[->] (-0.2,0) -- node[above, lbl, rotate=90] {query blocks} (-0.2,-\s*8);
  \node[lbl, anchor=west, align=left] (unit) at (\s*8+0.35,-\s*5.5)
    {selection unit: the query rows\\that share one selection decision};
  \draw[lead] (\s*6,-\s*5.5) -- (unit.west);
  \node[lbl, anchor=west, align=left] (diag) at (\s*8+0.35,-\s*2.5)
    {diagonal block:\\always kept};
  \draw[lead] (\s*3,-\s*2.5) -- (diag.west);
  \node[lbl, anchor=west, align=left, text=sysink] at (\s*8+0.35,-\s*7.6)
    {\tikz\fill[sysblue!70] (0,0) rectangle (0.2,0.13);\ selected key block\quad
     \tikz\draw[sysgrid] (0,0) rectangle (0.2,0.13);\ skipped};
\end{tikzpicture}%

%% file: figures/wafer.tex
\begin{tikzpicture}[sys,
    mesh/.style={draw=sysline, fill=sysfill, minimum size=3.6mm, inner sep=0pt, line width=0.4pt},
    full/.style={draw=sysline, fill=sysblue!70, minimum width=1.6mm, minimum height=1.5mm, inner sep=0pt},
    free/.style={draw=sysline, fill=white, minimum width=1.6mm, minimum height=1.5mm, inner sep=0pt},
    port/.style={->, sysline, line width=0.5pt},
  ]
  \draw[kv] (0,1.5) -- (2.55,1.5);
  \foreach \x in {0,...,6} {
    \foreach \y in {0,...,2} {
      \node[mesh] at (0.5*\x, 0.5*\y) {};
    }
  }
  \node[store, minimum size=3.6mm] (src) at (0,1.5) {};
  \node[note, anchor=south, text=sysink] at ($(src.north)+(0,0.04)$) {source};
  \foreach \i/\f in {1/0,2/1,3/0,4/1} {
    \node[pe, minimum size=3.6mm] (c\i) at (0.5*\i,1.5) {};
    \node[full] at ($(c\i.north)+(-0.095,0.15)$) {};
    \ifnum\f=1
      \node[full] at ($(c\i.north)+(0.095,0.15)$) {};
    \else
      \node[free] at ($(c\i.north)+(0.095,0.15)$) {};
    \fi
    \node[note, anchor=south, text=sysink] at ($(c\i.north)+(0,0.25)$) {P\textsubscript{\i}};
  }
  \node[draw=sysbluedark, line width=0.8pt, minimum size=3.6mm, inner sep=0pt] (mh) at (c2) {};
  \node[lbl, anchor=north, align=center] at (1.5,-0.25)
    {(a) PE mesh (755 $\times$ 1,170 usable) with one\\multicast route: block \textit{j}+1 waits,
     as P\textsubscript{2}\\and P\textsubscript{4} have no free receive buffer};
  \node[note, anchor=north, text=sysink] at (1.5,-1.12)
    {\tikz\node[full]{};\,occupied\quad\tikz\node[free]{};\,free buffer};
  \begin{scope}[shift={(5.35,0.75)}]
    \node[box, fill=white, minimum width=2.75cm, minimum height=1.3cm] (pe) at (0,0) {};
    \node[draw=sysbluedark, fill=sysbluelight, minimum width=0.7cm, minimum height=0.42cm,
          inner sep=0pt, line width=0.5pt] (rt) at (0,0.3) {router};
    \node[draw=sysline, fill=white, minimum width=0.8cm, minimum height=0.5cm,
          inner sep=1pt] (core) at (-0.72,-0.33) {CE};
    \node[draw=sysline, fill=white, minimum width=0.8cm, minimum height=0.5cm,
          inner sep=1pt, align=center] (sram) at (0.72,-0.33) {SRAM\\48 KB};
    \draw[port] (rt.south west) -- (core.north);
    \draw[port, <->] (core.east) -- (sram.west);
    \draw[port] (-1.65,0.3) -- node[above, note] {W} (rt.west);
    \draw[port] (rt.east) -- node[above, note] {E} (1.65,0.3);
    \draw[port] (0,0.98) node[right, note] {N} -- (rt.north);
    \node[lbl, anchor=north] at (0,-0.7) {(b) PE: router, compute\\engine (CE) and SRAM};
  \end{scope}
  \draw[sysbluedark, densely dotted] (mh.north east) -- (pe.north west);
  \draw[sysbluedark, densely dotted] (mh.south east) -- (pe.south west);
\end{tikzpicture}%

%% file: sections/03_motivation.tex
\section{What limits sparse prefill on a wafer}
\label{sec:motivation}

We start from the conventional placement of query rows for distributed causal
attention, \emph{striped} placement, which gives token $t$ of a head to consumer
$t \bmod C$ of its $C$ consumers and balances causal work across processors~\cite{brandon2023striped}. We
make four observations, each measured on the WSE-3 simulator or estimated with the model
calibrated on it, that lead to the design of \autoref{sec:design}. Unless noted, the
masks are those of XAttention.

\begin{figure}[t]
  \centering
  \includegraphics{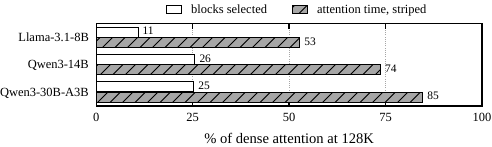}
  \caption{With striped placement, sparse attention still takes most of the dense
  attention time. Mask density and attention time of striped sparse prefill at 128K,
  relative to dense attention, over all layers (model).}
  \label{fig:striped-attention}
\end{figure}

\paragraph*{O1: With striped placement, the speedup is far below the reduction in work}
At 128K tokens, the masks select 11--26\% of the causal blocks. Yet with
striped placement, the attention stage still takes 53--85\% of the dense attention time
(\autoref{fig:striped-attention}). With selection and the query permutation included, end-to-end prefill is
then only 1.02--1.35$\times$ faster than dense prefill (\autoref{fig:sweep}). Striped
sparse prefill breaks even with dense prefill only at 58K--123K tokens with XAttention,
and only at 218K with MInference on Qwen3-30B-A3B.
On the wafer, the pruned work thus does not set the speedup; O2 and O3 show that the
placement of rows does (\autoref{sec:design}).

\begin{figure}[t]
  \centering
  \includegraphics{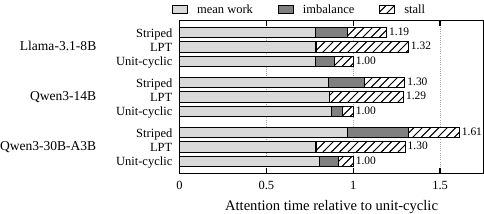}
  \caption{Balancing the total work leaves the stall on the shared stream. Attention
  time of the slowest KV column under each placement, three layers at 128K, summed,
  relative to unit-cyclic;
  imbalance: the slowest consumer's work above the mean (model).}
  \label{fig:column-parts}
\end{figure}

\paragraph*{O2: The bottleneck is stall time on shared streams, not data volume or total work}
If the placement decides the speedup, what should it optimize? Data volume is not the
answer. Delivering a 64-key block costs 1,056 cycles and serves every PE on the route, while
computing one selected (row, block) pair costs 2,365 cycles on one PE. If every group
had its own KV source, each column of \autoref{fig:column-parts} would take as long as
its slowest consumer, because the data transfer overlaps with compute. Total work is not the answer either.
Longest-processing-time-first (LPT) scheduling~\cite{graham1969bounds} keeps the load
imbalance of every column within 4.2\%. Yet its layers are
1.29--1.32$\times$ slower than with unit-cyclic placement (\autoref{sec:eval-placement}).
\autoref{fig:column-parts} shows why: its groups stall on the shared stream
1.8--2.3$\times$ as long as with striped placement.

The remaining cause is backpressure. Consider $R$ consumers that share one
route, each with $B$ receive buffers. They form a fork-join system with finite
buffers~\cite{baccelli1992synchronization}. Block $j$ takes $\tau$ cycles to deliver,
and consumer $r$ spends $w_{rj}$ cycles on it, with $w_{rj}=0$ if its rows skip the
block. Block $j$ can enter the route only when every consumer has finished block $j-B$:
\begin{equation}
\begin{aligned}
d_j &= \max\bigl(d_{j-1},\ \max_{r} f_{r,j-B}\bigr) + \tau,\\
f_{rj} &= \max\bigl(f_{r,j-1},\ d_j\bigr) + w_{rj},
\end{aligned}
\label{eq:stream}
\end{equation}
where $d_j$ is the delivery time of block $j$ and $f_{rj}$ the time at which consumer
$r$ finishes it. \autoref{eq:stream} is a max-plus
recurrence~\cite{baccelli1992synchronization}. Its makespan is the weight of the
maximum-weight path through the $(r,j)$ grid. The path can switch consumers, and each
switch skips $B-1$ blocks. Write $W_r(j)=\sum_{i\le j} w_{ri}$ for the cumulative work of
consumer $r$, and suppose that all consumers have equal totals. The stall is then at
least the \emph{prefix imbalance} $\Lambda_B=\max_{j,r,r'} W_{r'}(j)-W_r(j+B-1)$, the
largest lead of one consumer's cumulative work over another's $B-1$ blocks later. For
$N$ blocks, the stall is at most $\lfloor (N-1)/B\rfloor\Lambda_B$.
\autoref{fig:intro}(a) shows four such consumers.

On the simulator, \autoref{eq:stream} predicts the time of groups and columns within a
few percent (\autoref{sec:eval-setup}). With contiguous placement, which gives each
consumer a range of consecutive tokens, a group of 64 consumers
is 2.73$\times$ slower than the same consumers alone.
In short, the route advances at the rate of its busiest consumer. A placement must
therefore equalize the consumers' cumulative work at every point of the stream, not only
at its end (\S\S\ref{sec:design-group} and~\ref{sec:design-column}).

\begin{figure}[t]
  \centering
  \includegraphics{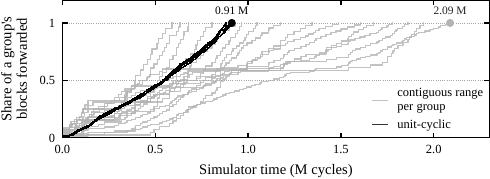}
  \caption{With unit-cyclic placement, the groups that share a KV column advance in
  lockstep. Share of its selected blocks that each of 16 groups of 16 consumers has
  forwarded on one KV column of 128 blocks; markers: the last group's last block
  (Llama-3.1-8B masks, simulator).}
  \label{fig:column-progress}
\end{figure}

\paragraph*{O3: The rows of one selection unit have identical work}
O2 shows that equal cumulative work at every point of the stream removes the stall.
All rows of a unit select the same blocks, so outside the diagonal block they add the
same work to every block of the stream. If every consumer of a group holds the same
number of rows of every unit, all consumers have the same work on every block, and
then the group runs as fast as one consumer:
\begin{proposition}
\label{prop:equal}
In \autoref{eq:stream}, if $w_{rj}=w_j$ for every consumer $r$ and block $j$, every
consumer finishes every block when one consumer alone, with the same $B$, $\tau$ and
$w_j$, would. The group thus takes as long as one consumer, for any $B\ge1$ and
$\tau\ge0$.
\end{proposition}
\begin{proof}
By induction on $j$: if $f_{ri}=f_i$ for every $r$ and $i<j$, then
$d_j=\max(d_{j-1},f_{j-B})+\tau$ and $f_{rj}=\max(f_{j-1},d_j)+w_j$. These are the
recurrence of one consumer, and both are independent of $r$.
\end{proof}
\noindent\autoref{fig:intro}(b) shows four such consumers.

The same argument applies one level up, to the groups that share a KV column. We ran
16 groups of 16 consumers that share one column on the simulator, with Llama-3.1-8B
masks (\autoref{fig:column-progress}). When the rows of every unit are spread
cyclically over the column, the groups advance in lockstep and all finish by 0.91\,M
cycles. When each group holds a contiguous range of units, they finish between
0.59\,M and 2.09\,M cycles.

Equal work needs every unit to have a multiple of $G$ rows in the group. Methods that
select per token and share the selection across a KV head, such as NSA and DeepSeek
sparse attention, have units of four rows. On the simulator, such groups stall by 0--33\%, against
13--173\% with contiguous placement (\autoref{sec:design-group}). Splitting units along
another axis also breaks equal work: giving each consumer the rows of only one query head is
1.7--2.1$\times$ slower.
The unit of placement should therefore be the selection unit. Because the placement
depends only on the shape of the units, it can be fixed offline
(\S\S\ref{sec:design-group} and~\ref{sec:design-policy}).

\begin{figure}[t]
  \centering
  \includegraphics{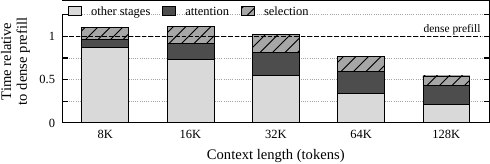}
  \caption{Up to 32K, selection adds more time than sparse attention saves. Sparse
  prefill of Llama-3.1-8B with unit-cyclic placement by stage, relative to dense
  prefill; other stages: GEMMs and the query permutation (model).}
  \label{fig:prefill-stages}
\end{figure}

\paragraph*{O4: Selection, not placement, sets the break-even length}
With rows placed by unit, the cost of the selection estimate decides, through the
break-even length, whether sparse prefill is faster than dense prefill. The estimate is a dense pass over the keys at stride 8. At 8K--32K
tokens, it costs more than the attention it saves. For Llama-3.1-8B at 8K, selection
adds 14\% of the dense prefill, and sparse attention saves only 4\%
(\autoref{fig:prefill-stages}). At 128K, selection adds 11\%, and sparse attention saves
57\%.
Selection must therefore run efficiently on the attention PEs, and the system
must choose dense or sparse prefill for each length
(\S\S\ref{sec:design-selection} and~\ref{sec:design-policy}).

%% file: sections/04_design.tex
\section{\sys{} design}
\label{sec:design}

\begin{figure*}[t]
  \centering
  \resizebox{\textwidth}{!}{\input{figures/overview}}
  \caption{\sys{} overview: a placement fixed offline from tensor shapes gives every
  consumer of a multicast group the same work on every KV block. Steps \step{1}--\step{5}
  execute at run time on every KV column; colored bars: selection units.}
  \label{fig:overview}
\end{figure*}
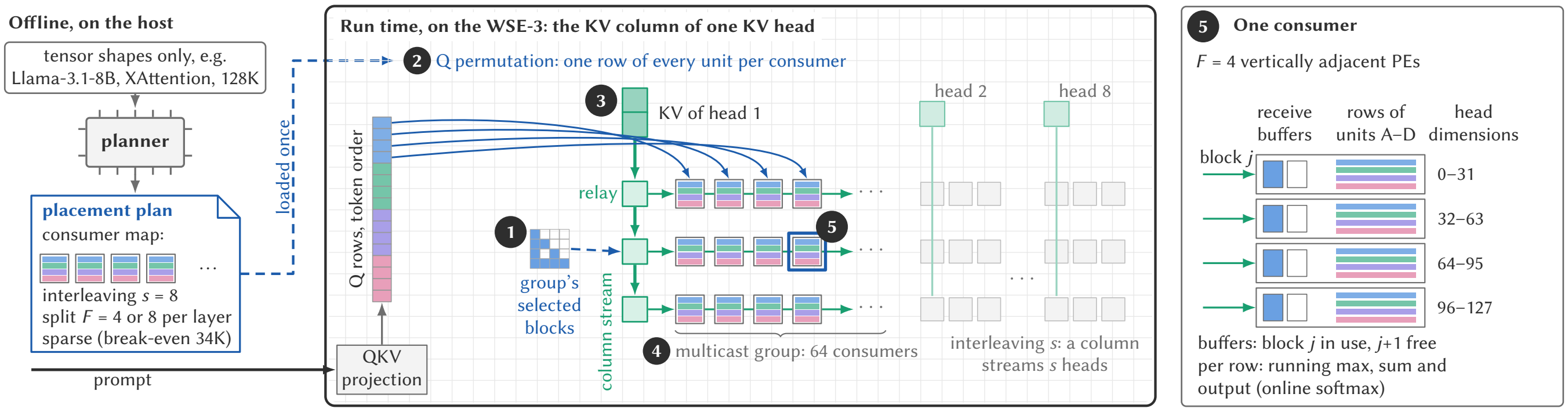

\sys{} places query rows so that the consumers of every stream have equal cumulative
work at every point of it (O2), with the selection unit as the unit of placement (O3).
Its unit-cyclic placement has two levels, a cyclic order inside each multicast group
(\autoref{sec:design-group}) and a per-head start offset across the groups of a KV column
(\autoref{sec:design-column}); \sys{} also interleaves KV heads across columns
(\autoref{sec:design-heads}). Each stream carries only the selected blocks
(\autoref{sec:design-stream}). \sys{} runs selection on the attention PEs (O4,
\autoref{sec:design-selection}), moves rows with one permutation
(\autoref{sec:design-placement}) and fixes every choice offline
(\autoref{sec:design-policy}). It accepts any dynamic block-sparse method and takes its
masks unchanged, because it needs only the shape of the method's selection unit: 128
tokens of one head (XAttention), 64 tokens of one head (MInference) or one token of
every query head of a KV head (NSA, DeepSeek sparse attention).

Take one layer of Llama-3.1-8B at 128K tokens (\autoref{fig:overview}), with a method
that selects per 128-token query block and head, such as XAttention, so a selection unit
is 128 consecutive tokens of one query head. The example first leaves out head
interleaving, which \autoref{sec:design-heads} adds: each of the layer's eight KV heads
runs on its own KV column of 27,136 consumers of $F=4$ PEs, in 424 multicast groups of
$G=64$, with one relay per group (a PE in each of its $F$ PE rows) and one KV source per
column. Offline, the planner on
the host computes the placement plan from tensor shapes: the consumer of every query row,
the interleaving degree, the head-dimension split $F$, and whether the length runs
sparse or dense prefill (\autoref{sec:design-policy}).

At run time, the QKV projection, WaferLLM's distributed GEMM
(MeshGEMM)~\cite{he2025waferllm}, leaves the query rows in token order. The attention
PEs then select every unit's blocks \step{1} (\autoref{sec:design-selection}), so each
relay knows the blocks its group selected and each consumer those of each of its rows.
One permutation moves every query row to its consumer \step{2}
(\autoref{sec:design-placement}), which then holds 19 or 20 rows, one per unit of its
group.

The KV source sends down the column stream \step{3} every block that at least one group
selected, and every relay receives each block into one of its five receive buffers. Each
relay forwards on its group's multicast route only the blocks that a row of the group
selected \step{4}; the route advances only when all 64 consumers have a free receive
buffer, so one busy consumer stalls its whole group (O2). For each block it receives, a
consumer computes the rows that selected it with an online softmax \step{5}, and it
receives block $j+1$ in its second buffer while it computes block $j$. Because a
consumer's work on a block depends on the rows it holds, step \step{2} decides whether
step \step{4} stalls. The inverse permutation returns the outputs to token order for the
output projection and the FFN, which for a mixture-of-experts model includes the router,
the dispatch and the experts, sized by measured load.

\subsection{Unit-cyclic placement inside a group}
\label{sec:design-group}

\begin{figure}[t]
  \centering
  \resizebox{\columnwidth}{!}{\input{figures/levels}}
  \caption{Unit-cyclic placement (a) inside a multicast group and (b) across the groups
  of a KV column, and (c) head interleaving across columns. A consumer's colored bars are
  the selection units it holds rows of; in (c), the KV heads.}
  \label{fig:levels}
\end{figure}
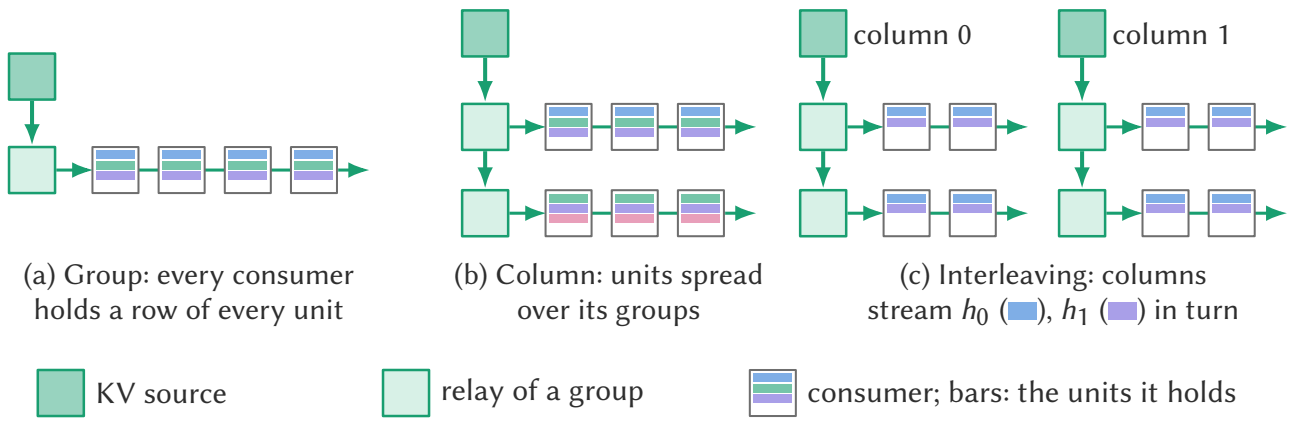

O3 shows that the rows of one selection unit have identical work on every block of
the stream. Inside each multicast group of $G$ consumers, \sys{} therefore lists the
rows of each unit one after another and gives row $i$ of each unit to consumer $i \bmod G$
(\autoref{fig:levels}(a)). This is a cyclic distribution~\cite{blackford1997scalapack}
of each unit's rows. Whenever every unit has a multiple of $G$ rows in the group, each
consumer then holds the same number of rows of every unit. In the example, a group of 64 consumers holds 64 consecutive
tokens of each unit it serves, one token per consumer, so every consumer holds one row
of every unit of the group. The rows of a unit differ only inside the diagonal block,
where causal masking gives each row a different number of keys. Outside that block, all
consumers have the same work on every block, so by \autoref{prop:equal} the group takes
nearly as long as one consumer alone, for any buffer depth.

\paragraph*{Other unit shapes}
The rule needs only the number of rows of each unit. MInference selects per 64-token
query block of one head, so with $G=64$ every consumer again holds one row of every unit
of its group. NSA and DeepSeek sparse attention share one selection among the query
heads of a KV head and select per token, so a unit has one row per query head, four for
Llama-3.1-8B. \sys{} lists these units in token order over the whole column. Such a unit
has fewer rows than a group has consumers, so different consumers hold rows of different
units, and their work along the stream differs.
On the simulator, these groups stall by 0--33\%, against 13--173\% with contiguous
placement (O3). Smaller groups reduce the stall, but at 128K, groups of 16--32
consumers need 1.4--4.4\% more PEs than the WSE-3 offers.

\paragraph*{Correctness}
Placement changes only where a row runs: each row keeps its token index and its
selected blocks, so every placement computes the same attention outputs, which we check
against FP64 (\autoref{sec:impl}); \autoref{sec:eval-placement} compares eight other
placements.

\subsection{Across the groups of a KV column}
\label{sec:design-column}

O2 shows that PEs that share a stream stall each other when their work falls at
different points of it, and the groups of a KV column share one. Each relay receives
the column stream into a few receive buffers (five at $F=4$ and ten at $F=8$, at 128K)
and acts as a consumer of the column: its work on a block is the time its group takes
to accept the block. \autoref{eq:stream} thus applies to the relays of a column as it
applies to the consumers of a group, and \sys{} also spreads every head's units over the
groups of the column.

Keys stream in token order, and a row at token $t$ works only on blocks up to $t$, so a
group's work along the stream depends on the token positions it holds. Dealing every
head from consumer 0 would give each group the same few positions of every head. \sys{}
therefore gives each query head a start offset~\cite{blackford1997scalapack}: with $M$
groups and $H$ query heads, the rows of head $k$ start at
the first consumer of group $\lfloor kM/H\rfloor$, so token $t$ of head $k$ (both from 0)
goes to
consumer $(t + G\lfloor kM/H\rfloor) \bmod MG$. Starting each head at a group boundary
keeps every 64-token block of a head on one group. In the example, the four query heads
start a quarter of the column apart, at groups 0, 106, 212 and 318. Each group then
holds 19 or 20 chunks of 64 consecutive tokens, four or five per head, with each head's
chunks 6,784 tokens apart from the previous head's. Every group thus receives units from
every head and every part of the sequence (\autoref{fig:levels}(b)), so its cumulative
work stays close to the column's mean. When the consumers do not divide a head's rows,
their row counts differ by at most one row per head.

For per-head units, striped placement is this rule with every head starting at consumer
0. Each group then holds the same positions of every head, so its work follows only four
or five positions, and the groups that hold late tokens, whose rows select more blocks, fall
behind. We evaluate this level by measurement: on the simulator at 32K, it makes a
column 1.16$\times$ faster than striped placement, and in the performance model at 128K,
a random start per head is 5--15\% slower than the even offsets
(\autoref{sec:eval-placement}).

\subsection{Across columns: head interleaving}
\label{sec:design-heads}

Each column has its own stream, so across columns only the total work matters: the next stage
of the layer starts when the slowest column finishes. KV heads select different numbers
of blocks, so the columns of a layer finish at different times, and \sys{} therefore
interleaves the heads over the columns. With interleaving of degree $s$, a set of $s$ KV heads shares $s$ columns. Each
column streams the blocks of all $s$ heads in round-robin order, with block $b$ of head
$h$ at position $sb+h$. The rule of \autoref{sec:design-column} then applies over the $sMG$
consumers and $sH$ query heads of the $s$ columns (\autoref{fig:levels}(c)). With $s=2$, for example, each
of two columns streams the blocks of both heads in turn and holds rows of both. At 128K
with XAttention, the planner picks $s=8$ for Llama-3.1-8B and Qwen3-14B, so every column
streams the blocks of all eight KV heads, and $s=4$ for Qwen3-30B-A3B, which has four KV
heads. Interleaving averages the heads'
work over the columns, but it makes the stream of every column $s$ times longer. It is
faster when the heads' work differs more than the longer stream costs.
The planner chooses $s\in\{1,2,4,8\}$ per model, method and length
(\S\S\ref{sec:design-policy} and~\ref{sec:eval-breakdown}).

\begin{figure}[t]
  \centering
  \resizebox{\columnwidth}{!}{\input{figures/streams}}
  \caption{Each stream carries only the blocks that its consumers selected (schematic, 8
  key blocks). The relays OR their groups' unit bitmaps into group bitmaps, and the KV
  source ORs these into the column bitmap. Relay 2 receives blocks 2 and 4 but does not forward
  them; a consumer of group 1 computes its A row on blocks 1 and 4.}
  \label{fig:streams}
\end{figure}
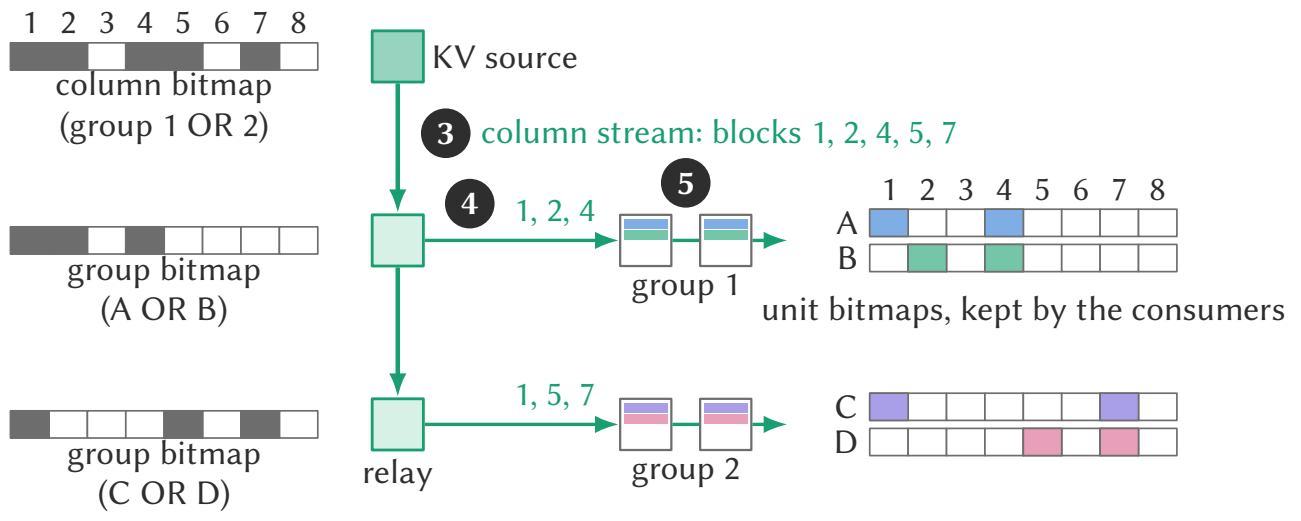

\subsection{Executing a block-sparse stream}
\label{sec:design-stream}

The placement turns pruned work into saved time only if every stream carries the
selected blocks alone and every consumer computes only the rows that selected them. \sys{}
therefore turns each unit's mask into bitmaps before attention (\autoref{fig:streams}).

\paragraph*{Block bitmaps}
Selection leaves one bitmap per unit, one bit per 128-key block for XAttention (1,024
bits at 128K); the streams carry 64-key blocks, so each such bit covers two of them.
\sys{} sends each unit's bitmap to the groups that hold the unit's rows, where each
consumer keeps the bitmaps of its rows. The bitmaps are ORed along each group's PE row
into the relay's group bitmap, and the group bitmaps along the KV column into the KV
source's column bitmap. The column stream \step{3} thus carries only blocks that some
group selected, and each group's route \step{4} only blocks that some row of the group
selected. For each block of the column stream, a relay reads its group bitmap: it forwards a selected block on its
group's route and frees the buffer of any other block, as relay 2 does for blocks 2 and
4 in \autoref{fig:streams}.

\paragraph*{Attention on a consumer}
A consumer \step{5} splits its rows over its $F$ PEs by head dimension, $128/F$
dimensions per PE. For each block it receives, every PE computes the partial scores of the rows whose
bitmap selects the block, and the $F$ PEs sum them with an all-reduce along their column.
Each PE then updates the online-softmax state of those rows (running maximum and sum)
and accumulates the weighted values of its own dimensions. On the diagonal block, a causal mask removes the keys after each row.

\paragraph*{PE and SRAM use}
The KV columns of all heads tile the wafer: at 128K with $F=4$, the attention of one
layer uses 99.8\% of the usable PEs, 98\% of them in consumers and the rest in relays
and KV sources. A consumer PE's code, query rows, buffers, bitmaps and softmax state take
34--39\,KB of its 48\,KB of SRAM; each further receive buffer would cost 2--4\,KB.
The layer's stages run one after another on the same PEs, and the
projections and the FFN use WaferLLM's GEMM grids (validated in
\autoref{sec:eval-setup}).

\subsection{Selection on the attention PEs}
\label{sec:design-selection}

Because selection sets the break-even length (O4), we run it as its own phase \step{1} on
the attention PEs, where the masks are used, before the attention stream. A selection
estimate scores query rows against every key block, so it streams the head's K blocks
once, densely, down the same KV column, and every consumer receives every block. We
implement the estimates of two published methods. For XAttention, every consumer holds
the strided query rows of two units, one early and one late in the sequence, so all
consumers score about the same number of (row, block) pairs on the shared stream. The
scores of a unit are summed over its PEs and consumers into one score per 128-key
block, in FP32, and one PE per unit selects the blocks with the published threshold
rule; the bitmaps go to the relays and consumers (\autoref{sec:design-stream}). For
MInference, a head's PEs score its last 64 queries and build the vertical and slash
indices.

\paragraph*{Selection by bisection}
Both rules keep the top of a ranking: XAttention keeps the highest-scoring blocks of a
unit until their share of the unit's total score reaches the threshold, and MInference
keeps the $k$ highest-scoring vertical and slash lines of a head. \sys{} finds the cut
by bisection instead of sorting. The scores are non-negative, so their FP32 bit patterns
order like integers, and a bisection on the bit pattern of the cut value takes at most
32 steps. For XAttention, the selecting PE makes one summing pass over the unit's block
scores per step and one more pass to add the next block below the cut, as the published
rule does;
for MInference, each step counts locally and sums the counts along the PE row. On the
simulator, XAttention's selection equals the reference in every compared block, and
\autoref{sec:eval-boundary} measures the cost of selection.

\subsection{Query permutation}
\label{sec:design-placement}

Unit-cyclic placement needs the query rows on their consumers, but the projection
leaves them in token order: it stores each chunk of 32,768 query rows along the PE rows
of its GEMM grid. \sys{} therefore moves every query row to its consumer with one
permutation \step{2} over the on-chip network, and the outputs return by the inverse
permutation. Because the placement ignores the masks, this permutation is the same for
every prompt of a given length and is part of the plan. We model each direction as the load of its busiest link plus its longest
path, with link costs calibrated on the simulator. At 128K tokens, both directions
together cost 1.5\,M cycles per layer, 2.7\% of a Llama-3.1-8B layer.

\subsection{Offline planner}
\label{sec:design-policy}

O3 shows that the placement depends only on the shapes of the units, and O4 shows that
selection, which depends on the length, sets the break-even length. The planner
therefore fixes every choice before a prompt arrives; at run time, only the block bitmaps
change with the prompt. Per model, method and length, it
chooses the interleaving degree $s$, the head-dimension split $F\in\{4,8\}$ of each
layer's heads, and whether to run sparse or dense prefill: below the break-even length
of \autoref{sec:eval-boundary}, the selection costs more than it saves, and the system
runs dense attention. Doubling $F$ halves the consumers of a KV column and doubles
their rows (at 128K, 13,568 consumers of up to 39 rows at $F=8$); the planner keeps, per
layer, the split that is faster in the performance model. For a mixture-of-experts model, the planner also places
the experts. Like DeepSeek-V3~\cite{deepseekai2024deepseek}, it adds redundant experts
by load: each expert gets square sub-grids of the projection grids, and each spare one
goes to the expert with the most tokens per sub-grid in a calibration window.

%% file: figures/overview.tex
\begin{tikzpicture}[sys,
    mpe/.style={draw=sysline!70, fill=white, minimum width=2.1mm, minimum height=1.3mm,
                inner sep=0pt, line width=0.3pt},
    faint/.style={draw=sysline!45, fill=sysfill, minimum size=2.6mm, inner sep=0pt, line width=0.3pt},
    perm/.style={->, sysbluedark, line width=0.55pt, >={Latex[length=1.2mm,width=0.9mm]}},
  ]
  \node[head, anchor=north west] at (0,6.55) {Offline, on the host};
  \node[box, fill=white, align=center, font=\sffamily\figsmall, text=sysink, inner sep=2pt,
        minimum width=2.35cm] (shapes) at (1.47,5.92) {tensor shapes only, e.g.\\Llama-3.1-8B, XAttention, 128K};
  \draw[->, sysline, line width=0.6pt] (shapes.south) -- (1.47,5.46);
  \node[draw=sysline, fill=sysfill, minimum width=1.1cm, minimum height=0.55cm, inner sep=0pt,
        line width=0.6pt, rounded corners=1pt] (cpu) at (1.47,5.08) {};
  \foreach \d in {-0.4,-0.2,0,0.2,0.4} {
    \draw[sysline, line width=0.6pt] ($(cpu.north)+(\d,0)$) -- ++(0,0.08);
    \draw[sysline, line width=0.6pt] ($(cpu.south)+(\d,0)$) -- ++(0,-0.08);
  }
  \foreach \d in {-0.15,0.15} {
    \draw[sysline, line width=0.6pt] ($(cpu.west)+(0,\d)$) -- ++(-0.08,0);
    \draw[sysline, line width=0.6pt] ($(cpu.east)+(0,\d)$) -- ++(0.08,0);
  }
  \node[font=\sffamily\bfseries\figsmall] at (cpu) {planner};
  \draw[->, sysline, line width=0.6pt] (1.47,4.72) -- (1.47,4.5);
  \draw[sysbluedark, fill=white, line width=0.6pt]
    (0.3,2.72) -- (0.3,4.48) -- (2.4,4.48) -- (2.65,4.23) -- (2.65,2.72) -- cycle;
  \draw[sysbluedark, line width=0.6pt] (2.4,4.48) -- (2.4,4.23) -- (2.65,4.23);
  \node[head, text=sysbluedark, anchor=north west] at (0.38,4.44) {placement plan};
  \node[note, text=sysink, anchor=west] at (0.38,4.0) {consumer map:};
  \foreach \k in {0,...,3} { \unitpe{pm\k}{0.56+0.4*\k,3.66}{unitA,unitB,unitC,unitD} }
  \node[note, text=sysink, anchor=west] at (2.15,3.66) {\dots};
  \foreach \t [count=\i] in {interleaving \textit{s} = 8, split \textit{F} = 4 or 8 per layer,
                             sparse (break-even 34K)} {
    \node[note, text=sysink, anchor=west] at (0.38,{3.58-0.24*\i}) {\t};
  }
  \draw[ctrl, line width=0.8pt] (2.65,3.6) -- (3.25,3.6) -- (3.25,6.0) -- (4.4,6.0);
  \node[note, rotate=90, anchor=south, text=sysbluedark] at (3.25,4.8) {loaded once};
  \node[draw=sysink, line width=0.8pt, rounded corners=3pt, minimum width=9.35cm,
        minimum height=4.5cm, anchor=south west] (wafer) at (3.6,2.1) {};
  \begin{scope}
    \clip[rounded corners=3pt] ($(wafer.south west)+(0.05,0.05)$) rectangle
                               ($(wafer.north east)+(-0.05,-0.05)$);
    \draw[sysgrid!80, line width=0.25pt, step=0.24] (3.6,2.09) grid (13.0,6.62);
  \end{scope}
  \node[head, anchor=north west, fill=white, inner sep=2pt] at ($(wafer.north west)+(0.1,-0.08)$)
    {Run time, on the WSE-3: the KV column of one KV head};
  \node[draw=sysline, fill=sysfill, align=center, font=\sffamily\figsmall, text=sysink,
        inner sep=1.5pt, line width=0.4pt] (proj) at (4.25,2.52) {QKV\\projection};
  \draw[flow] (0.3,2.52) -- node[below, note, text=sysink, pos=0.3] {prompt} (proj.west);
  \foreach \u/\c [count=\b from 0] in {A/unitA,B/unitB,C/unitC,D/unitD} {
    \foreach \k in {0,...,3} {
      \pgfmathsetmacro{\yy}{5.3-0.13*(4*\b+\k)}
      \node[mpe, fill=\c] (q\u\k) at (4.25,\yy) {};
    }
  }
  \node[note, rotate=90, anchor=south, text=sysink] at (4.1,4.33) {Q rows, token order};
  \draw[->, sysline, line width=0.6pt] (proj.north) -- (4.25,3.28);
  \node[store, minimum size=2.8mm] (st1) at (7.1,5.55) {};
  \node[store, minimum size=2.8mm] (st2) at (7.1,5.28) {};
  \node[note, anchor=west, text=sysink] at (7.32,5.42) {KV of head 1};
  \node[num] at (6.72,5.55) {3};
  \foreach \g/\y in {1/4.5, 2/3.85, 3/3.2} {
    \node[relay, minimum size=2.8mm] (r\g) at (7.1,\y) {};
    \foreach \k in {1,...,4} { \unitpe{c\g\k}{7.28+0.44*\k,\y}{unitA,unitB,unitC,unitD} }
    \draw[kv, line width=0.6pt] (r\g.east) -- (c\g1.west);
    \foreach \a/\b in {1/2,2/3,3/4} { \draw[systeal, line width=0.6pt] (c\g\a.east) -- (c\g\b.west); }
    \draw[kv, line width=0.6pt] (c\g4.east) -- ++(0.36,0);
    \node[text=sysline] at (9.78,\y) {$\cdots$};
  }
  \draw[kv, line width=1pt] (st2.south) -- (r1.north);
  \draw[kv, line width=1pt] (r1.south) -- (r2.north);
  \draw[kv, line width=1pt] (r2.south) -- (r3.north);
  \node[note, rotate=90, anchor=south, text=systeal] at (6.88,2.95) {column stream};
  \node[note, anchor=east, text=systeal] at (r1.west) {relay};
  \draw[decorate, decoration={brace, amplitude=2pt, mirror}, sysline]
    ($(c31.south west)+(0,-0.07)$) -- ($(c34.south east)+(0.72,-0.07)$)
    node[midway, below=1pt, note] {\step{4}\ multicast group: 64 consumers};
  \begin{scope}[shift={(5.92,4.1)}]
    \foreach \r in {0,...,3} { \foreach \c in {0,...,3} {
      \pgfmathtruncatemacro{\sel}{(\c<=\r && (\c==0 || \c==\r || \r==3)) ? 1 : 0}
      \ifnum\sel=1 \fill[sysblue!70] (0.11*\c,-0.11*\r) rectangle ++(0.11,-0.11); \fi
      \draw[sysline!60, line width=0.2pt] (0.11*\c,-0.11*\r) rectangle ++(0.11,-0.11);
    } }
  \end{scope}
  \node[num] at (5.7,4.07) {1};
  \node[note, anchor=north, align=center, text=sysbluedark] at (6.14,3.63) {group's\\selected\\blocks};
  \draw[ctrl] (6.38,3.88) -- (r2.west);
  \foreach \k/\t in {0/1,1/2,2/3,3/4} {
    \draw[perm] (qA\k.east) .. controls ($(qA\k.east)+(1.1,0.1)$) and ($(c1\t.north)+(-0.3,0.6)$) .. (c1\t.north);
  }
  \node[note, anchor=west, text=sysbluedark] at (4.45,6.0) {\step{2}\ Q permutation: one row of every unit per consumer};
  \foreach \x/\lab in {10.45/2, 11.85/8} {
    \node[draw=systeal!60, fill=systeal!20, minimum size=2.8mm, inner sep=0pt] at (\x,5.4) {};
    \foreach \y in {4.5,3.85,3.2} {
      \node[faint] at (\x,\y) {};
      \foreach \k in {1,2} { \node[faint] at (\x+0.32*\k,\y) {}; }
    }
    \draw[systeal!45, line width=0.6pt] (\x,5.25) -- (\x,3.35);
    \node[note, anchor=south] at (\x+0.32,5.55) {head \lab};
  }
  \node[text=sysline] at (11.48,3.52) {$\cdots$};
  \node[note, align=center, anchor=north] at (11.72,2.95) {interleaving \textit{s}: a column\\streams \textit{s} heads};
  \draw[sysbluedark, line width=1pt] ($(c24.south west)+(-0.04,-0.04)$) rectangle ($(c24.north east)+(0.04,0.04)$);
  \node[num, anchor=south west] at ($(c24.north east)+(-0.02,-0.02)$) {5};
  \node[box, fill=white, minimum width=4.3cm, minimum height=4.5cm, anchor=south west]
    (zoom) at (13.25,2.1) {};
  \node[num, anchor=north west] at ($(zoom.north west)+(0.12,-0.1)$) {5};
  \node[head, anchor=north west] at ($(zoom.north west)+(0.55,-0.1)$) {One consumer};
  \node[note, anchor=north west, text=sysink] at ($(zoom.north west)+(0.12,-0.5)$)
    {\textit{F} = 4 vertically adjacent PEs};
  \node[note, text=sysink, align=center, anchor=south] at (14.42,5.06) {receive\\buffers};
  \node[note, text=sysink, align=center, anchor=south] at (15.45,5.06) {rows of\\units A--D};
  \node[note, text=sysink, align=center, anchor=south] at (16.55,5.06) {head\\dimensions};
  \foreach \f/\y/\lab in {0/4.72/{0--31}, 1/4.22/{32--63}, 2/3.72/{64--95},
                          3/3.22/{96--127}} {
    \draw[kv, line width=0.6pt] (13.5,\y) -- (14.1,\y);
    \node[draw=sysline, fill=white, minimum width=1.9cm, minimum height=0.44cm, inner sep=0pt,
          line width=0.4pt] at (15.05,\y) {};
    \fill[sysblue!70, draw=sysline, line width=0.3pt] (14.18,\y-0.15) rectangle (14.4,\y+0.15);
    \draw[sysline, fill=white, line width=0.3pt] (14.45,\y-0.15) rectangle (14.67,\y+0.15);
    \foreach \c [count=\i from 0] in {unitA,unitB,unitC,unitD} {
      \fill[\c] (15.0,{\y+0.165-0.09*\i}) rectangle (15.9,{\y+0.105-0.09*\i});
    }
    \node[note, anchor=west, text=sysink] at (16.1,\y) {\lab};
  }
  \node[note, anchor=south west, text=sysink] at (13.42,4.76) {block \textit{j}};
  \node[note, anchor=north west, align=left, text=sysink] at (13.4,2.95)
    {buffers: block \textit{j} in use, \textit{j}+1 free\\per row: running max, sum and\\output (online softmax)};
\end{tikzpicture}%

%% file: figures/levels.tex
\begin{tikzpicture}[sys]
  \newcommand{\grouprow}[2]{
    \draw[kv, line width=0.6pt] (#1.east) -- ++(0.26,0);
    \draw[systeal, line width=0.6pt] ($(#1.east)+(0.26,0)$) -- ++({0.46*(#2-1)+0.32},0);
    \draw[kv, line width=0.6pt] ($(#1.east)+({0.46*(#2-1)+0.58},0)$) -- ++(0.22,0);
  }
  \node[store] (as) at (0,0.95) {};
  \node[relay] (ar) at (0,0.3) {};
  \draw[kv] (as.south) -- (ar.north);
  \grouprow{ar}{4}
  \foreach \k in {1,...,4} { \unitpe{ap\k}{0.12+0.46*\k,0.3}{unitA,unitB,unitC} }
  \node[lbl, anchor=north] at (1.09,-0.3) {(a) Group: every consumer\\holds a row of every unit};
  \begin{scope}[shift={(3.15,0)}]
    \node[store] (bs) at (0,1.25) {};
    \foreach \g/\y/\ua/\ub/\uc in {1/0.6/unitA/unitB/unitC, 2/0/unitB/unitC/unitD} {
      \node[relay] (br\g) at (0,\y) {};
      \grouprow{br\g}{3}
      \foreach \k in {1,2,3} { \unitpe{bq\g\k}{0.12+0.46*\k,\y}{\ua,\ub,\uc} }
    }
    \draw[kv] (bs.south) -- (br1.north);
    \draw[kv] (br1.south) -- (br2.north);
    \node[lbl, anchor=north] at (0.86,-0.3) {(b) Column: units spread\\over its groups};
  \end{scope}
  \begin{scope}[shift={(5.5,0)}]
    \foreach \h/\x in {0/0, 1/1.8} {
      \node[store] (ct\h) at (\x,1.25) {};
      \node[anchor=west, note, text=sysink] at (ct\h.east) {column \h};
      \foreach \g/\y in {1/0.6, 2/0} {
        \node[relay] (cr\h\g) at (\x,\y) {};
        \grouprow{cr\h\g}{2}
        \foreach \k in {1,2} { \unitpe{cw\h\g\k}{\x+0.12+0.46*\k,\y}{unitA,unitC} }
      }
      \draw[kv] (ct\h.south) -- (cr\h1.north);
      \draw[kv] (cr\h1.south) -- (cr\h2.north);
    }
    \node[lbl, anchor=north] at (1.6,-0.3) {(c) Interleaving: columns\\stream
      \textit{h}\textsubscript{0} (\tikz\fill[unitA] (0,0) rectangle (0.2,0.12);),
      \textit{h}\textsubscript{1} (\tikz\fill[unitC] (0,0) rectangle (0.2,0.12);) in turn};
  \end{scope}
  \begin{scope}[shift={(0.2,-1.25)}]
    \node[store] at (0,0) {};
    \node[note, anchor=west, text=sysink] at (0.2,0) {KV source};
    \node[relay] at (2.4,0) {};
    \node[note, anchor=west, text=sysink] at (2.6,0) {relay of a group};
    \unitpe{lg}{4.95,0}{unitA,unitB,unitC}
    \node[note, anchor=west, text=sysink] at (5.15,0) {consumer; bars: the units it holds};
  \end{scope}
\end{tikzpicture}%

%% file: figures/streams.tex
\begin{tikzpicture}[sys]
  \def\cw{0.24}
  \def\ch{0.17}
  \def\lx{-2.42}
  \def\rx{2.95}
  \newcommand{\streambits}[4]{%
    \foreach \b in {1,...,8} {
      \draw[sysline, line width=0.3pt, fill=white]
        ({#1+(\b-1)*\cw},{#2-\ch/2}) rectangle ++(\cw,\ch);
    }
    \foreach \b in {#4} {
      \draw[sysline, line width=0.3pt, fill=#3]
        ({#1+(\b-1)*\cw},{#2-\ch/2}) rectangle ++(\cw,\ch);
    }
  }
  \newcommand{\streamdigits}[2]{%
    \foreach \b in {1,...,8} { \node[note, text=sysink] at ({#1+(\b-0.5)*\cw},#2) {\b}; }
  }
  \newcommand{\streamrow}[3]{%
    \draw[kv, line width=0.6pt] (#1.east) -- (#2.west);
    \draw[systeal, line width=0.6pt] (#2.east) -- (#3.west);
    \draw[kv, line width=0.6pt] (#3.east) -- ++(0.22,0);
  }
  \node[store] (s) at (0,2.45) {};
  \node[relay] (r1) at (0,1.3) {};
  \node[relay] (r2) at (0,0.15) {};
  \draw[kv] (s.south) -- (r1.north);
  \draw[kv] (r1.south) -- (r2.north);
  \node[note, anchor=west, text=sysink] at (s.east) {KV source};
  \node[note, text=sysink] at (0,-0.17) {relay};
  \node[num, minimum size=3.1mm] at (0.3,1.97) {3};
  \node[note, anchor=west, text=systeal] at (0.48,1.97) {column stream: blocks 1, 2, 4, 5, 7};
  \streamdigits{\lx}{2.66}
  \streambits{\lx}{2.45}{sysink!70}{1,2,4,5,7}
  \node[note, text=sysink] at (\lx+4*\cw,2.13) {column bitmap\\(group 1 OR 2)};
  \streambits{\lx}{1.3}{sysink!70}{1,2,4}
  \node[note, text=sysink] at (\lx+4*\cw,0.98) {group bitmap\\(A OR B)};
  \streambits{\lx}{0.15}{sysink!70}{1,5,7}
  \node[note, text=sysink] at (\lx+4*\cw,-0.17) {group bitmap\\(C OR D)};
  \unitpe{c11}{1.55,1.3}{unitA,unitB}
  \unitpe{c12}{2.05,1.3}{unitA,unitB}
  \streamrow{r1}{c11}{c12}
  \node[num, minimum size=3.1mm] at (0.45,1.53) {4};
  \node[note, text=systeal] at (0.98,1.47) {1, 2, 4};
  \node[num, minimum size=3.1mm] at (1.8,1.66) {5};
  \node[note, text=sysink] at (1.8,1.0) {group 1};
  \node[note, text=sysink] at (1.8,-0.15) {group 2};
  \unitpe{c21}{1.55,0.15}{unitC,unitD}
  \unitpe{c22}{2.05,0.15}{unitC,unitD}
  \streamrow{r2}{c21}{c22}
  \node[note, text=systeal] at (0.98,0.32) {1, 5, 7};
  \streamdigits{\rx}{1.62}
  \foreach \u/\y in {A/1.41, B/1.19, C/0.26, D/0.04} {
    \node[note, text=sysink, anchor=east] at (\rx-0.03,\y) {\u};
  }
  \streambits{\rx}{1.41}{unitA}{1,4}
  \streambits{\rx}{1.19}{unitB}{2,4}
  \streambits{\rx}{0.26}{unitC}{1,7}
  \streambits{\rx}{0.04}{unitD}{5,7}
  \node[note, text=sysink] at (\rx+4*\cw,0.85) {unit bitmaps, kept by the consumers};
\end{tikzpicture}%

%% file: sections/05_implementation.tex
\section{Implementation}
\label{sec:impl}

We implement \sys{} on the WSE-3 with Cerebras SDK 2.10.1. Its wafer program gives
each PE one of three roles, KV source, relay or consumer, and runs the XAttention or
MInference selection phase on the consumers; one program runs a whole attention layer,
selection, forwarding and attention, on the SDK's WSE-3 simulator. The PE programs are written in CSL, the SDK's language, about 17,000 lines with
experimental variants. The XAttention selection phase streams K in a dimension-major
order, and a fused scoring loop cuts its time to 0.854 of an unfused version. The planner, the performance model and
the experiment scripts are about 37,000 lines of Python; the host writes the planner's
output into each consumer's SRAM through the SDK's memcpy interface. We capture masks
and run RULER with the published XAttention, MInference and RULER code, unchanged
(PyTorch 2.11, Transformers 5.14).

We check the outputs of every run: attention must be within 2\% relative $L_2$ error of
an FP64 reference, and XAttention selection must equal a NumPy transcription of the
published code, which matches the published PyTorch code on 32,400 (head, query block,
key block) entries. MInference's selection phase matches the reference on 38,523 of
38,524 indices, the one miss a near-tie. In the layer program, on Llama-3.1-8B queries
and keys at 4K tokens, all 2,248 selected blocks match and outputs are within 0.44\% of
FP64.

%% file: sections/06_evaluation.tex
\section{Evaluation}
\label{sec:eval}

We answer four research questions:
\begin{itemize}
\item \textbf{RQ1} (\autoref{sec:eval-e2e}): How much faster is \sys{} than dense and
  striped sparse prefill, and from which context length?
\item \textbf{RQ2} (\autoref{sec:eval-breakdown}): Which mechanism gives the speedup, and how
  does unit-cyclic placement compare with other placements and mask-aware load
  balancers?
\item \textbf{RQ3} (\autoref{sec:eval-robust}): Does the speedup hold with a faster
  attention kernel, other network parameters, balancing by unit at run time and
  measured expert loads?
\item \textbf{RQ4} (\autoref{sec:eval-quality}): Does sparse prefill keep the model's
  accuracy?
\end{itemize}

\subsection{Experimental setup}
\label{sec:eval-setup}

\paragraph*{Performance model}
A 128K layer needs the whole wafer and does not fit the simulator, so we estimate
end-to-end prefill with an analytical model calibrated on simulator measurements. Per
layer, it applies \autoref{eq:stream} to every column and group, with the consumers'
per-row and per-block costs and the link costs measured on the simulator, and adds the
selection stage, the query permutation, the GEMM and elementwise stages (fitted on
WaferLLM's MeshGEMM stages) and, for Qwen3-30B-A3B, the MoE router, dispatch and expert
FFN with the experts each token selects in a captured CPU prefill. It runs a layer's
stages in sequence, assumes that weight loads and KV writes overlap
compute, and omits the LM head (\autoref{sec:discussion}).

We multiply the attention time by a calibration factor of 1.08--1.125, the measured
ratio of the attention phase to the model's in small simulator runs, and report the
lower end; the two ends differ by at most 0.03 in speedup. We report cycles and ratios,
because the WSE-3 clock is not published. All speedups are of end-to-end prefill, and K
denotes 1,024 tokens.

\paragraph*{Model validation}
On held-out simulator runs, the model's mean error per component is 1.2--4.9\%, from
one consumer on a stream to multicast groups, groups that share a column, selection and
MeshGEMM steps, and its largest single error is 10.9\%, for one consumer at low density.
It predicts whole attention layers at 8K (Llama-3.1-8B layers 16 and 24, one feature
slice per consumer, 16,648 PEs) within 1.3\%, naming the
slowest column in every run, and a 32K column of 8,192 consumers under three placements
within 1.6\%, in the measured order. From 8K to 32K, the model overestimates every
measured speedup between two placements, interleaving degrees or masks by at most
1.6\%, and dense and sparse masks give the same error on an 8K column. In the layer
program, consumers take 1.125 and 1.156 times the model's attention time at 32K and
64K; 1.156 would raise the speedups by at most 2.7\%.

\paragraph*{Workloads and baselines}
We use three instruction-tuned models: Llama-3.1-8B~\cite{llamateam2024llama},
Qwen3-14B (with YaRN scaling~\cite{peng2024yarn} by 4 above 32K) and Qwen3-30B-A3B
(release 2507~\cite{qwenteam2025qwen3}; 128 experts, 8 active per token, up to 256K). We capture XAttention masks on all three and MInference masks on
Llama-3.1-8B and Qwen3-30B-A3B, from full-model prefill on one prompt per length, the
first tokens of the WikiText-2 validation text~\cite{merity2016pointer}, on RTX A6000
GPUs or
CPUs, which only capture masks and run RULER; every reported time comes from the
simulator or the model. Unless
noted, results use XAttention masks. All
baselines run on the same simulated wafer with the same masks, kernels, GEMM stages and
costs; only the placement differs. Dense prefill uses the same mapping with every causal
block; its kernel runs at about half of the peak multiply-add rate, and against a dense
kernel at the peak rate \sys{} still achieves a 1.24--1.78$\times$ speedup without
expert routing (\autoref{sec:eval-robust}). Striped sparse prefill, the fastest of
four conventional placements (\autoref{sec:eval-placement}), places token $t$ on
consumer $t \bmod C$ and is tuned over the same head-dimension splits as \sys{}.

\paragraph*{Quality}
We run RULER's~\cite{hsieh2024ruler} 13 tasks with greedy decoding and its metrics,
10 samples per task and length (at 128K, 50 and 30 for dense and stride 8 on
Llama-3.1-8B and Qwen3-14B), and report each method's score minus dense prefill's on the
same prompts.

\subsection{RQ1: Speedup and break-even length}
\label{sec:eval-e2e}

\sys{} is 1.31--1.49$\times$ faster than striped sparse prefill at each model's longest
context. Sparse prefill thus becomes 1.47--1.90$\times$ faster than dense prefill
instead of 1.13--1.35$\times$. \sys{} also breaks even with dense prefill at
1.4--2.7$\times$ shorter contexts.

\begin{figure*}[t]
  \centering
  \includegraphics{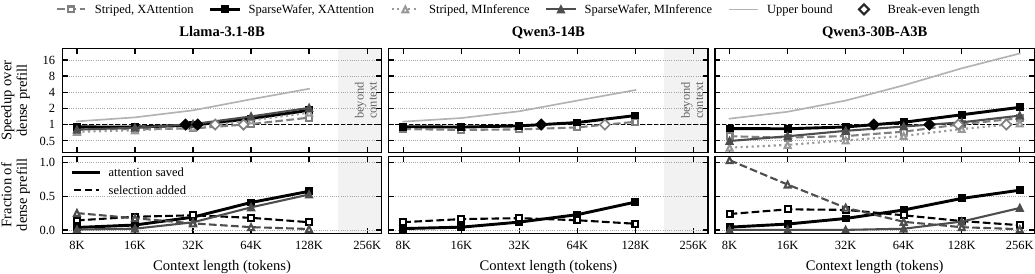}
  \caption{\sys{} breaks even with dense prefill at 29--87K tokens, 1.4--2.7$\times$
  shorter than striped sparse prefill (42--218K). Top: end-to-end speedup over dense
  prefill; the upper bound sets sparse attention and selection to zero cost. Bottom:
  attention time saved and selection time added, as fractions of dense prefill (model).}
  \label{fig:sweep}
\end{figure*}


\paragraph*{Speedup at long contexts}
The gain comes from attention: at 128K, the attention stage of \sys{} takes 27--48\% of
the dense attention time, against 53--85\% with striped placement
(\autoref{fig:striped-attention}). Reducing the stall thus turns more of the pruned work
into saved time (O1). The speedup follows the mask density: at 128K, Llama-3.1-8B's
masks have a density of 11\% and give 1.85$\times$, and Qwen3-14B's have 26\% and give
1.47$\times$. On three more 128K WikiText-2 prompts, these speedups change by at most
1.4\%.
\autoref{fig:sweep} leaves out expert routing for Qwen3-30B-A3B
(\autoref{sec:eval-moe}); without it, \sys{} is 1.47--2.10$\times$ faster than dense
prefill with XAttention and 1.23--1.59$\times$ faster than striped sparse prefill with
either method.

\paragraph*{Speedup across context lengths}
With XAttention, the gain over striped placement grows with context length, because
attention time grows
quadratically and takes a larger share of the prefill. For Llama-3.1-8B, it rises from
1.09$\times$ at 8K to 1.37$\times$ at 128K.

\paragraph*{Break-even length}
\label{sec:eval-boundary}
\sys{} breaks even with dense prefill at 29--87K tokens with either method (34--45K
with XAttention), against 42--218K for striped placement (the diamonds in
\autoref{fig:sweep}), so sparse prefill is faster than dense prefill for a wider range of
prompt lengths.
The bottom row of \autoref{fig:sweep} explains this length. The speedup is
$1/(1-a+e+p)$, where $a$ is the attention time saved, $e$ the time that selection adds
and $p$ the permutation time added over dense prefill's, each as a fraction of the dense
prefill. The term $p$ stays at or below 1\%, so the break-even length is where the
saved attention overtakes selection (O4). Saved attention grows with length, because
causal work grows quadratically and the density falls: for Llama-3.1-8B, the density
drops from 32\% at 8K to 11\% at 128K, and the saved attention rises from 4\% to 57\% of
the dense prefill. MInference on Qwen3-30B-A3B breaks even last, at 87K, because its
masks have a density of 100\% at 8K and 16K and its index building adds 1.03 of the dense
prefill at 8K. Below the break-even length, the planner runs dense prefill
(\autoref{sec:design-policy}), so \sys{} is at least as fast as dense prefill at every
length.

\paragraph*{Selection speed}
The fused scoring loop (\autoref{sec:impl}) shortens every XAttention break-even length
by 8--14\%. With the FP32 accumulation that exact selection needs, it takes 0.71 cycles
per multiply-add, against 0.5 at the peak multiply-add rate, and three alternative
layouts gain at most another 15\% in the performance model. At 128K, selection takes
13--21\% of the sparse prefill; stride 16 would cut it by 36\% but loses accuracy
(\autoref{sec:eval-quality}).

\paragraph*{Distance to the upper bound}
The upper bound in \autoref{fig:sweep} is the speedup that Amdahl's law allows if sparse
attention and selection take zero time. \sys{} reaches 40\% of it on Llama-3.1-8B and
34\% on Qwen3-14B at 128K. Qwen3-30B-A3B activates 8 of 128 experts, so its projections
and FFN take a small share of the prefill and its bound is high; \sys{} reaches 10\% of
it at 256K. Besides selection, the remaining time is attention, which takes 27--48\% of
the dense attention time for masks that select 11--26\% of the blocks. This ratio is
about the product of five factors of 1.1--1.4 each: (1) the chosen head-dimension split,
1.12--1.36; (2) per-block costs on every 64-key block that any of a consumer's rows
selected, 1.09--1.13; (3) imbalance between columns, 1.13--1.19; (4) imbalance inside a
column, 1.12--1.20; and (5) stall on the shared stream, 1.19--1.24.

\takeaway{\sys{}'s speedup over dense prefill grows with context length, and it breaks
even at 1.4--2.7$\times$ shorter contexts than striped placement, because equal work
along each shared stream turns pruned attention into saved time.}

\subsection{RQ2: Where the speedup comes from}
\label{sec:eval-breakdown}

Unit-cyclic placement gives most of the gain on the Qwen3 models and about half on
Llama-3.1-8B, and it is faster than every other offline placement and every mask-aware
load balancer; only balancing by unit at zero decision cost is faster
(\autoref{sec:eval-robust}).

\begin{figure}[t]
  \centering
  \includegraphics{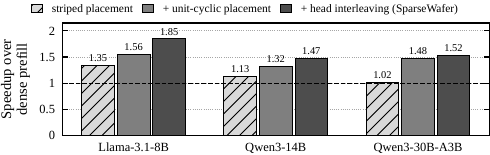}
  \caption{Unit-cyclic placement gives most of the gain over striped placement on the
  Qwen3 models and about half on Llama-3.1-8B; interleaving adds the rest. At 128K
  (model).}
  \label{fig:steps}
\end{figure}

\paragraph*{Breakdown by mechanism}
\autoref{fig:steps} adds the mechanisms of \sys{} one at a time, at 128K and with every
other cost fixed. Unit-cyclic placement over the groups of
each KV column, without head interleaving, makes end-to-end prefill 1.16--1.45$\times$ faster
than striped placement. Striped groups also meet \autoref{prop:equal}, so this gain
comes from the per-head start offset, which keeps the cumulative work of every group
close to the column's mean (\autoref{sec:design-column}); striped groups stall on the
column stream (\autoref{fig:column-parts}). Unit-cyclic placement gains most
on Qwen3-30B-A3B, where striped sparse prefill is only 1.02$\times$ faster than dense
prefill.

\begin{figure}[t]
  \centering
  \includegraphics{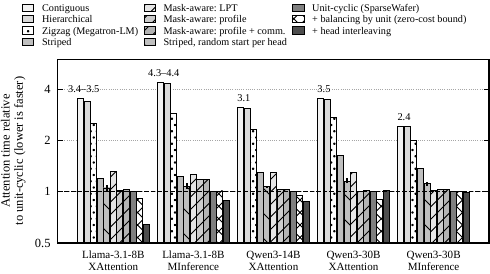}
  \caption{Unit-cyclic placement is faster than every conventional placement and every
  mask-aware load balancer, even with the balancers' decisions at zero cost: 0.5--17\%
  faster than the best balancer per case. Attention time of three layers at 128K
  (geometric mean; Qwen3-30B: Qwen3-30B-A3B). Balancing by unit: 64--80\,K run-time
  decisions (\autoref{sec:eval-robust}); random start: mean, range of five seeds;
  $s=8,8,8,4,2$ (model).}
  \label{fig:placement}
\end{figure}

\paragraph*{Head interleaving}
Head interleaving then adds 1.12--1.19$\times$ on Llama-3.1-8B and Qwen3-14B, but only
1.03$\times$ on Qwen3-30B-A3B. On the three layers per model of \autoref{fig:placement},
it adds 1.13--1.55$\times$ on Llama-3.1-8B and Qwen3-14B, and 0.98--1.01$\times$ on
Qwen3-30B-A3B; the planner picks $s$ from all layers of each model. Interleaving
is faster only when the KV heads' work differs more than its longer stream costs
(\autoref{sec:design-heads}). On the simulator's whole layer (\autoref{sec:eval-setup}),
interleaving is 1.20--1.27$\times$ faster than without it, within 1.6\% of the
performance model; on layer 24, both give 0.99$\times$. Interleaving is thus a gain that
depends on the LLM.

\paragraph*{Placements compared}
\label{sec:eval-placement}
We compare nine placements on three layers per model and method at 128K
(\autoref{fig:placement}), each with its faster head-dimension split and one KV column
per head. Contiguous placement gives each consumer a range of tokens, and hierarchical
placement gives each group a range, striped inside the group. Zigzag placement, the causal load balancing of Megatron-LM's context
parallelism~\cite{nvidia2026context}, cuts the sequence into $2C$ chunks and gives
consumer $c$ chunks $c$ and $2C-1-c$. Striped placement distributes tokens cyclically.
Three mask-aware load balancers read each layer's masks and place rows one at a time,
in decreasing order of work: longest processing time first
(LPT)~\cite{graham1969bounds} on the consumer with the least total work; the
profile-aware balancer on the group and consumer whose load profile along the stream
overlaps least with the row's; and the profile- and communication-aware balancer, which
adds the cost of blocks new to a group's bitmap, a greedy hypergraph-partitioning
heuristic~\cite{catalyurek1999hypergraph} whose weight we tune over nine values on a
held-out layer. \emph{Balancing by unit} starts from unit-cyclic placement and moves
whole units between groups by their measured work, after selection.

\paragraph*{Comparison with conventional placements}
Unit-cyclic placement is 2.4--4.4$\times$ faster than contiguous and hierarchical
placement. Their groups hold whole token ranges, so each group receives its work at a
different point of the stream (O2). Zigzag placement balances each consumer's causal
work, yet unit-cyclic placement is still 2.0--2.9$\times$ faster: a zigzag group holds
two short token ranges, so its work also falls at a few points of the stream.
Unit-cyclic placement is also 1.19--1.62$\times$ faster than striped placement; on the
simulator, at 32K, the slowest column measures 1.16$\times$ (modeled 1.17$\times$).
Striped groups do not stall inside, but they hold the same positions of every head, so
their work falls at different points of the column stream (\autoref{sec:design-column}).
The evenly spaced starts matter. With a
random start per head, attention is 5--15\% slower than with unit-cyclic placement on
average over five seeds (0.7--20\% per seed), and with adjacent starts it stays within
5\% of striped placement.

\paragraph*{Comparison with mask-aware load balancers}
Unit-cyclic placement is also faster than every mask-aware balancer on the mean of every
model and method, although the balancers read every mask and make 4.2--5.2\,M decisions
per layer. LPT is
the slowest of the three in four of the five cases, 1.02--1.32$\times$ slower than
unit-cyclic placement: it balances the totals but spreads the rows of one unit over
different groups, so every group's set of selected blocks grows and its groups stall
on the shared stream the longest (\autoref{fig:column-parts}). The profile-aware
balancers come within 0.5--4\% on XAttention, because they track the work along the
stream, but with a greedy search over millions of rows for every input; placing rows by
unit tracks it without any search. Balancing by unit at zero decision cost would save another 5--10\%
on XAttention (\autoref{sec:eval-robust}).

\takeaway{Unit-cyclic placement gives about half or more of the gain over striped
placement and is faster than mask-aware balancers that make millions of decisions per
layer, because it equalizes the work along each shared stream from tensor shapes.}

\subsection{RQ3: Sensitivity}
\label{sec:eval-robust}

The speedup holds with a faster attention kernel, other buffer and link costs, and
measured expert loads, and balancing by unit at run time costs more than it saves.

\paragraph*{Sensitivity to the attention kernel}
Dense and sparse prefill use the same attention kernel, at about half of the peak rate
of 2 multiply-adds per cycle per PE. A faster kernel shortens attention on both sides
but not the projections or the selection, so the speedup shrinks.

\begin{figure}[t]
  \centering
  \includegraphics{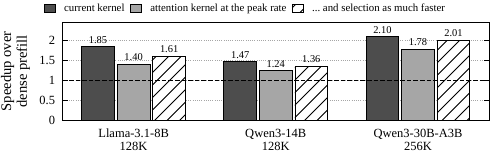}
  \caption{The speedup holds with a faster attention kernel. Speedup over dense prefill
  with the current attention kernel and with one 1.95--1.98$\times$ faster for dense and
  sparse prefill, at the peak multiply-add rate, with and without selection faster by the
  same factor (model; expert routing not included).}
  \label{fig:rate}
\end{figure}

At each model's longest length, we model a kernel that is 1.95--1.98$\times$ faster,
which brings dense attention to the peak multiply-add rate. \sys{} still achieves a
1.24--1.78$\times$ speedup over dense prefill there (\autoref{fig:rate}), and
1.36--2.01$\times$ if selection becomes faster by the same factor. The break-even length,
however, moves from 34--45K to 59--78K tokens if only the attention kernel improves, and
falls to 33--40K if selection improves with it. A faster attention kernel thus needs
faster selection at medium lengths (O4).

\paragraph*{Sensitivity to buffer, group and link costs}
The gain changes little with the buffer depth, the group size and the link cost up to 4
times the WSE-3's. We vary each one in the model at 128K, on three layers of
Llama-3.1-8B and Qwen3-30B-A3B with a head-dimension split of 4: 1--8
receive buffers, groups of 16--256 consumers, and link costs of 0.25--4 times the
WSE-3's. Over this range, unit-cyclic attention stays 1.17--1.19$\times$ (Llama-3.1-8B)
and 1.65--1.72$\times$ (Qwen3-30B-A3B) faster than striped attention. The gain
shrinks only where delivery, not compute, sets the time: at 16 times the link cost, it
is 1.01$\times$ on Llama-3.1-8B and 1.19$\times$ on Qwen3-30B-A3B. On the five cases of
\autoref{fig:placement}, a separate KV source for every group would remove the shared
stream and shorten unit-cyclic attention, without head interleaving, by at most 11\%.
Unit-cyclic placement on the shared stream is within 4\% of striped placement on that
idealized network for Llama-3.1-8B and 6--35\% faster for the other three cases, and LPT
on the idealized network would be 5--13\% faster than unit-cyclic placement. Equal work along the stream is thus the rule for
networks whose multicast is shared and flow-controlled.

\paragraph*{Cost of run-time balancing}
Balancing by unit (\autoref{fig:placement}, cross-hatched) costs more than it saves. It
needs a sequential decision after selection, on the critical path of each layer, since
each unit's group depends on the loads after the previous unit. With the SDK's measured
reductions, one decision takes about 3,500 cycles, so one column's decisions take
27--65\,M cycles per layer at 128K. That is 12--14 times the decision time below which
balancing would be faster for Llama-3.1-8B. Even a dedicated argmin path, which the
wafer does not have, would bring Qwen3-30B-A3B at 256K only to 1.92$\times$, against
1.90$\times$ for unit-cyclic placement. A parallel balancer decides quickly but is slower: sorting a column's units
by measured work and assigning them to the groups in serpentine order decides in
0.4\,M cycles per layer, but it equalizes the totals, not the work along the stream, so
attention becomes 16--26\% slower. End-to-end prefill then drops to
1.67$\times$ (Llama-3.1-8B, 128K) and 1.66$\times$ (Qwen3-30B-A3B, 256K), so
unit-cyclic placement is faster.

\begin{figure}[t]
  \centering
  \includegraphics{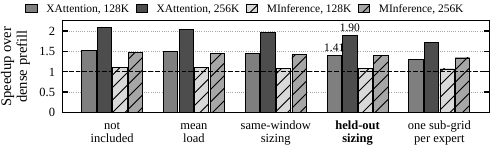}
  \caption{With measured expert loads, \sys{} stays faster than dense prefill on
  Qwen3-30B-A3B. Speedup with expert routing, per sizing of the redundant experts;
  held-out sizing (\sys) uses another window (model).}
  \label{fig:moe}
\end{figure}

\paragraph*{Measured expert loads}
\label{sec:eval-moe}
\autoref{fig:sweep} leaves out the MoE router and the dispatch of Qwen3-30B-A3B and gives
every expert the mean load. Real routing is skewed, so here we add the router, the
dispatch and the expert FFN with the experts that each token selected in a captured CPU
prefill, for dense and sparse
prefill alike (\autoref{fig:moe}). The planner sizes the redundant experts
(\autoref{sec:design-policy}) on one 32K-token window; we evaluate on another.

\begin{figure}[!tb]
  \centering
  \includegraphics{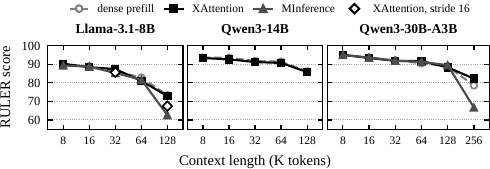}
  \caption{XAttention at stride 8 loses at most 1.8 points of RULER accuracy at every
  length. Mean of 13 tasks; 10 samples per task (at 128K, 50 and 30 for dense and stride
  8 on Llama-3.1-8B and Qwen3-14B); losses are on the same prompts (GPU measurements).}
  \label{fig:ruler}
\end{figure}

With measured, held-out expert loads, \sys{} on Qwen3-30B-A3B is 1.41$\times$ faster
than dense prefill at 128K and 1.90$\times$ at 256K. Measured routing lowers the speedup
by 7--10\%, because skewed loads lengthen the expert FFN for dense and
sparse prefill alike, so attention takes a smaller share. With one sub-grid per expert,
the speedup at 256K would fall from 2.04$\times$ at the mean load to 1.71$\times$;
redundant experts recover more than half of that loss. Sizing them on another window
costs 2--3\%, so it can be fixed offline.

\takeaway{The speedup of \sys{} holds with a faster attention kernel, other network
parameters and skewed expert loads, and its offline placement is faster than run-time
balancing by unit.}

\subsection{RQ4: Model quality}
\label{sec:eval-quality}

XAttention at stride 8 (RQ1--RQ3) loses at most 1.8 RULER points at any length, and
MInference 8.9--11.9 at each model's longest one.

\paragraph*{XAttention at stride 8}
On the same prompts, stride 8 scores at most 1.8 points below dense prefill at any
length and at most 0.6 below at 128K (\autoref{fig:ruler}); four cells score up to 3.7
points higher. The loss stays small because XAttention keeps, for every query block,
the key blocks with most of its estimated probability.

\paragraph*{MInference and stride 16}
MInference stays within 1.2 points of dense prefill up to 64K, and on Qwen3-30B-A3B
also at 128K. It picks one pattern of vertical and slash lines
per head from the head's last 64 queries, and at each model's longest length it falls
below dense prefill: by 8.9 points on Llama-3.1-8B at 128K and by 11.9 points on
Qwen3-30B-A3B at 256K. Stride 16 estimates each block from half the entries and is 6.5
points below stride 8 on Llama-3.1-8B at 128K.

\paragraph*{Speed on RULER masks}
RULER prompts produce denser masks than the WikiText-2 text of RQ1: XAttention's masks
are 1.25--1.63$\times$ denser on average, and denser on every task. On masks of one RULER
prompt per model with the task-average density, at 128K, \sys{} is
1.54$\times$ (Llama-3.1-8B) and 1.28$\times$ (Qwen3-14B) faster than dense prefill, against 1.85 and 1.47$\times$ on
WikiText-2. Its gain over striped placement holds at 1.39 and 1.27$\times$ (striped:
1.11 and 1.01$\times$ over dense): denser masks shrink the gain over dense prefill, not
the gain from placement.

\takeaway{\sys{} has the accuracy of its selection method, because it moves rows and
keeps masks: with XAttention at stride 8, it loses at most 1.8 points of RULER accuracy
at any length.}

%% file: sections/07_discussion.tex
\section{Discussion}
\label{sec:discussion}

\paragraph*{Generality and hardware implications}
The stall of \autoref{eq:stream} arises on any network whose multicast waits for every
destination, and unit-cyclic placement applies to such networks, because it needs only
the unit shapes and the group size; we evaluate it on one wafer to keep architecture and
mapping separate. On this network, hardware can still help: a KV
source per group would shorten unit-cyclic attention by up to 11\%, and deeper receive
buffers (2--4\,KB each) shorten every placement's stall.

\paragraph*{Scope of the evaluation}
The end-to-end results come from a performance model that predicts the simulator's
speedups within 1.6\%, validated up to a 64K column and whole 8K layers
(\autoref{sec:eval-setup}). It streams each layer's weights (0.44--1.25\,GB) in and its K
and V (0.54\,GB) out in parallel with compute. Even with these transfers serialized, at the CS-3's 150\,GB/s of
I/O~\cite{smith2026cerebras} and an assumed clock of 1.1\,GHz, the WSE-2's
maximum~\cite{he2025waferllm}, \sys{} is 1.43--1.84$\times$ faster than dense prefill. The
LM head adds under 0.6\%. We evaluate one prompt at a time; chunked and batched prefill
are future work.

%% file: sections/08_related.tex
\section{Related work}
\label{sec:related}

\paragraph*{Sparse and distributed attention}
XAttention~\cite{xu2025xattention}, MInference~\cite{jiang2024minference} and
FlexPrefill~\cite{lai2025flexprefill} select key blocks at run time to shorten prefill;
Quest~\cite{tang2024quest} selects pages for decoding, and NSA~\cite{yuan2025native} and
DeepSeek sparse attention~\cite{deepseekai2025deepseek} train the selection with the
model. They target GPUs, where kernels such as
FlashAttention~\cite{dao2022flashattention,dao2024flashattention} and
FlexAttention~\cite{dong2024flex} read only the selected blocks from memory. Across
devices, Ring Attention~\cite{liu2024ringattention} passes KV blocks around a ring,
Striped Attention~\cite{brandon2023striped} stripes tokens to balance causal work (our
main baseline), Megatron-LM's context parallelism splits the sequence in a
zigzag~\cite{nvidia2026context}, MTraining~\cite{li2025mtraining} stripes queries for
dynamic sparse attention in training, and SparSP~\cite{sun2026sparsp} cuts the traffic
of sequence parallelism. These methods balance total work or reduce traffic between
devices with independent memories. \sys{} takes the sparse methods' masks unchanged and
reduces the stall that these placements leave on a shared multicast stream
(\autoref{sec:eval-placement}).

\paragraph*{Spatial accelerators and work partitioning}
The Cerebras wafer-scale engine~\cite{lie2023cerebras,cerebras2026wafer} and
WaferLLM~\cite{he2025waferllm}, which maps dense LLM inference onto it, are the closest
prior work; we use WaferLLM's GEMM stages and add dynamic sparse attention. The SambaNova
SN40L~\cite{prabhakar2024sambanova}, the Graphcore IPU~\cite{jia2019dissecting}, Tesla
Dojo~\cite{talpes2023microarchitecture} and Tenstorrent
Blackhole~\cite{vasiljevic2024blackhole} also combine distributed SRAM with on-chip
networks. Quartz~\cite{golden2025quartz} spreads data
used together over all tiles of a sparse accelerator, with a partition computed from the
input, and SATA~\cite{fan2026sata} schedules selective token attention on one
accelerator. Our mask-aware baselines are greedy forms of LPT
scheduling~\cite{graham1969bounds}, Quartz's objective and hypergraph
partitioning~\cite{catalyurek1999hypergraph}. The stall
of \autoref{eq:stream} is a max-plus recurrence~\cite{baccelli1992synchronization}, and
its prefix-imbalance bound is a prefix discrepancy of the PEs' cumulative
work~\cite{bansal2022prefix}. These partitioners read the input; unit-cyclic placement
uses only shapes, makes the prefix imbalance zero in a group (\autoref{prop:equal}) and
keeps it small across groups with per-head start offsets (\autoref{sec:design-column}).

%% file: sections/09_conclusion.tex
\section{Conclusion}
\label{sec:conclusion}

On a statically routed wafer-scale accelerator, multicast backpressure between the PEs
that share a key-value stream removes most of the speedup of dynamic sparse prefill,
and balancing their total work leaves this stall in place. \sys{} distributes the query
rows of every selection unit cyclically over the PEs of each multicast group, so every
PE has the same work on every block and the group runs as fast as one PE; a per-head
start offset extends the placement across the groups that share a key-value source.
Fixed from tensor shapes, it takes any block-sparse method's masks unchanged and is
faster than mask-aware load balancers. On the WSE-3
simulator and a model calibrated on it, \sys{} with XAttention masks is
1.31--1.49$\times$ faster than striped placement at each model's longest context. Sparse
prefill thus becomes 1.47--1.90$\times$ faster than dense prefill instead of
1.13--1.35$\times$, and it loses at most 1.8 points of RULER accuracy.
On networks whose multicast waits for every destination, work that shares a stream must
be balanced at every point of the stream, and the shape of a sparse computation does
most of it.

%% file: sections/99_ack.tex
\begin{acks}
The authors used the AI assistants Claude (Anthropic) and ChatGPT (OpenAI) to help draft
and edit the text of this paper. The authors reviewed all text and take full responsibility
for the content.
\label{end-of-content}
\end{acks}